\documentclass[twocolumn, dvipsnames, twocolappendix]{aastex701}

\graphicspath{{./}{figures/}}

\usepackage{amsmath}	
\usepackage{graphicx}
\usepackage{xspace}
\usepackage{txfonts}
\usepackage{bm}

\usepackage{xpatch}
\usepackage{hyperref}

\hypersetup{
	colorlinks=true,
	breaklinks=true,
	citecolor=Blue,
	allcolors=Blue,
	frenchlinks=true
}

\makeatletter
\xpatchcmd\NAT@citex
{%
	\@citea\NAT@hyper@{%
		\NAT@nmfmt{\NAT@nm}%
		\hyper@natlinkbreak{\NAT@aysep\NAT@spacechar}{\@citeb\@extra@b@citeb}%
		\NAT@date
	}%
}
{%
	\@citea
	\NAT@nmfmt{\NAT@nm}%
	\NAT@aysep\NAT@spacechar
	\NAT@hyper@{\NAT@date}%
}
{}{}
\xpatchcmd\NAT@citex
{%
	\@citea\NAT@hyper@{%
		\NAT@nmfmt{\NAT@nm}%
		\hyper@natlinkbreak{\NAT@spacechar\NAT@@open\if*#1*\else#1\NAT@spacechar\fi}%
		{\@citeb\@extra@b@citeb}%
		\NAT@date
	}%
}
{
	\@citea
	\NAT@nmfmt{\NAT@nm}%
	\NAT@spacechar\NAT@@open\if*#1*\else#1\NAT@spacechar\fi
	\NAT@hyper@{\NAT@date}%
}
{}{}
\makeatother

\newcommand\citecrisp{Thomas et al. (in prep)}
\newcommand\citepcrisp{(Thomas et al. in prep, see also \citealt{thomas_why_2025})}

\defcitealias{thomas_why_2025}{T25}

\defcitealias{sike_cosmic-ray-driven_2025}{S25}
\newcommand\sttf{\citetalias{sike_cosmic-ray-driven_2025}\xspace}
\newcommand\sttfp{\citepalias{sike_cosmic-ray-driven_2025}\xspace}

\newcommand{\mhdX}{MHD\xspace}

\newcommand{\nlX}{CR-NL\xspace}
\newcommand{\inX}{CR-NL-IN\xspace}

\newcommand{\crmhd}{CRMHD\xspace}

\newcommand{\velsym}{\varv}

\begin{document}

\title{Resolving Star Cluster Formation in Galaxy Simulations with Cosmic Ray Feedback}

\correspondingauthor{Brandon Sike}
\email{bsike@umich.edu}

\author[0009-0008-1788-4355]{Brandon Sike}
\affiliation{Department of Astronomy, University of Michigan, Ann Arbor, MI 48109, USA}
\affiliation{Leibniz Institute for Astrophysics, Potsdam (AIP), An der Sternwarte 16, D-14482 Potsdam, Germany}
\email{bsike@umich.edu}

\author[0009-0002-2669-9908]{Mateusz Ruszkowski}
\affiliation{Department of Astronomy, University of Michigan, Ann Arbor, MI 48109, USA}
\email{mateuszr@umich.edu}

\author[0000-0001-9852-9954]{Oleg Y. Gnedin}
\affiliation{Department of Astronomy, University of Michigan, Ann Arbor, MI 48109, USA}
\email{ognedin@umich.edu}

\author[0000-0002-5970-2563]{Yingtian Chen}
\affiliation{Department of Astronomy, University of Michigan, Ann Arbor, MI 48109, USA}
\email{ybchen@umich.edu}

\author[0000-0002-2910-2276]{Matthias Weber}
\affiliation{Leibniz Institute for Astrophysics, Potsdam (AIP), An der Sternwarte 16, D-14482 Potsdam, Germany}
\email{maweber@aip.de}

\author[0000-0002-7443-8377]{Timon Thomas}
\affiliation{Leibniz Institute for Astrophysics, Potsdam (AIP), An der Sternwarte 16, D-14482 Potsdam, Germany}
\email{tthomas@aip.de}

\author[0000-0002-7275-3998]{Christoph Pfrommer}
\affiliation{Leibniz Institute for Astrophysics, Potsdam (AIP), An der Sternwarte 16, D-14482 Potsdam, Germany}
\email{cpfrommer@aip.de}

\begin{abstract}

Star clusters host the massive stars responsible for feedback in star-forming galaxies. Stellar feedback shapes the interstellar medium (ISM), affecting the formation of future star clusters. To self-consistently capture the interplay between feedback and star formation, a model must resolve the parsec-scale star formation sites and the multiphase ISM. Additionally, the dynamical impact of cosmic rays (CRs) on star formation rates (SFRs) must also be considered. We present the first simulations of the formation of an ensemble of star clusters with dynamically-coupled CRs, near-individual star particles, and a feedback-regulated ISM. We analyze tallbox simulations performed using the \textsc{Crisp} model in the moving-mesh code \textsc{Arepo}. We apply varied implementations of CR transport under the theory of self-confinement. We find that CRs simultaneously reduce the SFR, the power law slope of the cluster mass function, and the cluster formation efficiency. Each simulation is compatible with observations, and CR feedback tends to move results along observed star cluster relations. We see only modest changes in cluster radius and velocity dispersions, but significant differences in the virial parameters. Ultimately, the primary impact of CRs is to reduce SFRs. Lower SFRs imply fewer supernovae, and consequently a lower turbulent energy budget for gas. Star clusters formed in a CR-regulated ISM have lower velocity dispersions, and are therefore more bound under self-gravity. The effective clustering of supernovae is unchanged by CRs. Despite the idealized setup, the \textsc{Crisp} feedback model recovers many key aspects of star cluster formation.

\end{abstract}




\section{Introduction}
\label{sec:introduction}

\setcounter{footnote}{0}

Feedback in star-forming galaxies is overwhelmingly supplied by massive stars, which are primarily born in star clusters \citep{kroupa_inverse_1995, lada_embedded_2003}. Star cluster formation begins with the collapse of dense, filamentary molecular clouds \citep{bally_filamentary_1987, bergin_cold_2007, ballesteros-paredes_diffuse_2020} through hierarchical fragmentation \citep{bonnell_hierarchical_2003, vazquez-semadeni_global_2019}. Within the dense cores of these molecular clouds, star formation begins \citep{hartmann_rapid_2001, krumholz_star_2019, chevance_molecular_2020, krause_physics_2020, owen_secret_2023}, creating embedded clusters. This embedded stage is disrupted by a combination of radiation \citep{stromgren_physical_1939, oort_acceleration_1955, whitworth_erosion_1979}, winds \citep{castor_interstellar_1975, bally_protostellar_2016}, and supernovae \citep[SNe;][]{mckee_photoionized_1984, rogers_feedback_2013}. Further formation of stars is truncated by the disruption of the cloud, and the association of stars becomes an open cluster. A young star cluster can supply much more effective feedback than an individual isolated star, as consecutive SNe from clustered stars undergo less catastrophic cooling losses (due to ``preconditioning''), and therefore deliver more energy to the interstellar medium \citep[ISM;][]{mccray_supershells_1987, gatto_modelling_2015, naab_theoretical_2017}. A full characterization of galactic feedback must include an understanding of the formation, evolution, and disruption of star clusters, connecting these parsec-scale processes to the evolution of the galaxy as a whole.

Fortunately, there exists a wealth of observations of star clusters \citep{krumholz_star_2019, adamo_star_2020}, providing ample tests for theories of star cluster formation. The Gaia mission \citep{gaia_collaboration_gaia_2016} has supplied an unprecedented wealth of data on nearby stars used to characterize open clusters in the Milky Way galaxy \citep[e.g.,][]{almeida_open_2025}. Additionally, young star clusters in nearby galaxies have been characterized through their population-level statistics, including radii \citep{brown_radii_2021} and mass functions \citep[e.g.,][]{johnson_panchromatic_2017, messa_young_2018}. The mass functions of young star cluster populations appear to follow a power law of the approximate form $\mathrm{d}N/\mathrm{d}M\propto M^{\alpha_\mathrm{PL}}$, with $\alpha_\mathrm{PL} \approx -2$. However, the precise value of the power law slope varies with the environment, showing positive correlation with the surface density of the star formation rate  \citep[SFRD or $\Sigma_\mathrm{SFR}$;][]{adamo_star_2020}. 

Another population-level statistic is the cluster formation efficiency (CFE or $\Gamma$) -- the fraction of star formation that occurs in the form of bound star clusters \citep{bastian_star_2008}. The CFE encodes many processes during star cluster formation and evolution. The virial parameters (describing boundedness under self-gravity) of molecular clouds can vary within individual populations \citep{bertoldi_pressure-confined_1992, kauffmann_low_2013} and also depend on environment \citep{ni_life_2025, myers_gravitational_2025}. The boundedness of a giant molecular cloud (GMC) is then translated to the natal boundedness of the resulting star cluster \citep[e.g.,][]{kruijssen_fraction_2012}. Soon after these star clusters are born, they can be tidally disrupted by surrounding GMCs \citep{kruijssen_fraction_2012}, decreasing the CFE. By comparing the outputs of simulations with this wealth of observations, we can test our models and refine our theoretical understanding of galactic feedback.

Although cosmic rays (CRs) have received significant attention in the study of galactic feedback \citep{zweibel_basis_2017, ruszkowski_cosmic_2023, owen_cosmic_2023}, the impact of CRs on star cluster formation is uncertain. CRs are well-known for their contribution to heating within optically-thick molecular clouds \citep{padovani_impact_2020}. In addition to heating, CRs supply pressure to the ISM. The CR pressure in the solar neighborhood (and similar environments) is comparable to other sources of pressure \citep[e.g.,][]{boulares_galactic_1990, draine_physics_2011, crocker_cosmic_2021}, suggesting that CRs can be dynamically significant. Through their pressure, CRs can affect the landscape of the ISM, reducing SFRs \citep[e.g.,][]{salem_cosmic_2014, dashyan_cosmic_2020, kjellgren_dynamical_2025, sike_cosmic-ray-driven_2025} and driving galactic outflows \citep[e.g.,][]{socrates_eddington_2008, simpson_role_2016, rathjen_silcc_2023}. However, the impact of CRs depends on the adopted transport model \citep[e.g.,][]{uhlig_galactic_2012, pakmor_galactic_2016, ruszkowski_global_2017, farber_impact_2018, sike_cosmic-ray-driven_2025}. Indeed, accounting for uncertainties in transport, the true effect of CRs on the star-forming ISM is unknown. CRs may provide no vertical support for the weight of the ISM \citep{armillotta_cosmic-ray_2021}, CRs may not have any significant impact on SFRs \citep[e.g.,][]{buck_effects_2020, thomas_why_2025}, and CRs may have negligible dynamical impact on molecular cloud evolution \citep{fitz_axen_suppressed_2024, kjellgren_dynamical_2025}. In the pursuit of a comprehensive theory of galaxy formation, we must develop theoretically-motivated models for star cluster formation alongside the impact of CR feedback, and rigorously test these models against observations. Here, we discuss our efforts to model star cluster formation and CR feedback in recent simulations.

In the studies of star cluster formation, high-resolution simulations can now track star particles representative of individual stars \citep[e.g.,][]{hu_star_2016, wheeler_be_2019, calura_sub-parsec_2022}, allowing for the self-consistent clustering of SNe within the simulated ISM \citep[e.g.,][]{smith_efficient_2021, gutcke_lyra_2021,gutcke_lyra_2022-1, hu_code_2023, deng_rigel_2024}. These individual-star simulations allow for direct comparison to observed mass functions and CFEs \citep{li_disruption_2019, li_star_2019, lahen_griffin_2020, hislop_challenge_2022,gutcke_2024}. A promising alternative approach is to replace star clusters with subgrid sink particles \citep[e.g.,][]{li_star_2017, kim_three-phase_2017, rathjen_silcc_2021, reina-campos_star_2025}, overcoming the resolution requirements of simulating individual stars. These subgrid models can be tuned to match observations \citep[e.g.,][]{keller_empirically_2022} or tuned to match higher-resolution simulations \citep[e.g.,][]{li_star_2018, grudic_starforge_2021, grudic_great_2023, zhang_entangled_2025}. Notably, works adopting either approach (sink particle or individual star particles) must also account for the numerical impacts of different star formation prescriptions \citep{li_effects_2020, brown_testing_2022, hu_code_2023}.

In the studies of CR feedback, much work is being done to develop CR transport models motivated by relativistic plasma theory. For example, the theory of self-confinement \citep{kulsrud_effect_1969, skilling_cosmic_1975, shalaby_deciphering_2023, lemmerz_theory_2025} describes the propagation of CRs scattered by self-excited Alfvén waves. This theory has been applied with two-moment fluid models of CR transport \citep{jiang_new_2018, thomas_cosmic-ray_2019}, allowing for the self-consistent determination of CR propagation rates and momentum coupling to the gas. This approach allows for CR transport to vary with the underlying gas phase \citep[e.g.,][]{armillotta_cosmic-ray_2021, thomas_effective_2025, hix_dynamically_2025}, facilitating predictions for CR propagation that are consistent with observations \citep{chiu_simulating_2024, armillotta_energy-dependent_2025}. Another popular approach is to simply adopt a constant diffusion coefficient to match observations \citep[e.g.,][]{chan_cosmic_2019, wetzel_second_2025} and avoid the uncertainties surrounding the microphysics of CR transport. New tests for theoretical models are needed to constrain the real properties of CR transport, especially for theoretically-motivated models. If a simulation includes spatially-dependent transport (such as the theory of self-confinement), the simulation must include a resolved SN-heated hot phase \citep[e.g.,][]{armillotta_cosmic-ray_2024} and star-forming cold phase \citep[e.g.,][]{rathjen_silcc_2021, rathjen_silcc_2023} to self-consistently describe the impact of CRs on galaxies, especially pertaining to star formation.

In our previous work \citep[][hereafter \sttf]{sike_cosmic-ray-driven_2025}, we explored the impact of ion-neutral damping (IND) on  CR feedback. IND is a friction-like process that damps CR-excited Alfvén waves, reducing the momentum coupling between CRs and gas within mostly-neutral media \citep{kulsrud_effect_1969}. Previous work found that IND could reduce the impact of CRs on the ISM \citep[e.g.,][]{farber_impact_2018, armillotta_cosmic-ray_2021}. In \sttf, we found that IND does not prevent CRs from providing effective feedback; rather, CRs with IND lower SFRs by reducing the amount of star-forming gas in the ISM. Additionally, CRs with IND are able to drive a warm galactic wind with a moderate mass loading factor. The analysis of \sttf was primarily focused on CR feedback in the context of driving a galactic outflow. We presented bulk-scale properties of star formation such as the SFR, but did not investigate the impact of CR feedback on star formation in greater detail. Indeed, this is an aspect of CR feedback that is generally unexplored, especially with sophisticated CR transport models. 

For this work, we study the impact of CR feedback on star formation in the high-resolution ``tallbox'' simulations presented in \sttf. We found that these simulations produce clustering of star particles similar to observed stellar clusters. In this paper, we present the properties of these simulated star clusters, characterize their environmental dependence, and investigate the impact of CR feedback on star clusters. By comparing our simulated star clusters with observations of young star clusters, we can test the multiphase ISM model in our simulations, and provide additional constraints for CR transport models.

This paper is organized as follows. In Section~\ref{sec:methods}, we review our numerical model and the simulations presented in \sttf. Next, in Section~\ref{sec:star_particles_with_pretty_figure}, we look at the distribution of star particles in our fiducial model to motivate this work. Section~\ref{sec:mass_functions_section} presents the star cluster identification algorithm, the star cluster mass functions for the populations in the different cases, and the environmental dependencies of the mass functions. In Section~\ref{sec:sn_clustering}, we investigate the implications for the clustering of SNe. Next, in Section~\ref{sec:cluster_properties}, we investigate cluster properties such as radii, age distributions, velocity dispersion distributions, and virial parameters, followed by a discussion of the impact of CR feedback on star cluster formation in Section~\ref{sec:impact_of_crs}. Finally, we summarize and conclude this work in Section~\ref{sec:conclusions}.



\section{Simulation Methods}
\label{sec:methods}

For this work, we analyze the simulations presented in \sttf. Here, we review the aspects of the setup most relevant to this work.


\subsection{Overview}
\label{sec:simulation_overview}

The presented simulations were performed with the moving-mesh magnetohydrodynamics code \textsc{Arepo} \citep{springel_e_2010, pakmor_magnetohydrodynamics_2011, pakmor_simulations_2013, pakmor_improving_2016, weinberger_arepo_2020}. We also include self-gravity and (for two of the three cases) two-moment Alfvén wave-regulated \crmhd \citep{pfrommer_simulating_2017, thomas_cosmic-ray_2019, thomas_comparing_2022, thomas_finite_2021, thomas_cosmic-ray-driven_2023}. The thermal state of the ISM is evolved using the \textsc{Crisp} non-equilibrium thermochemistry and feedback model \citepcrisp.

Three primary (star-forming) simulations were presented in \sttf. These three simulations begin with an identical setup, but include different implementations of CR physics. The first case includes MHD, but treats cosmic rays as a uniform heating and ionization term \citep{glover_star_2007}. This first case is designated as the \mhdX case. The second case implements two-moment \crmhd with nonlinear Landau damping (NLLD) as the only Alfvén wave damping mechanism. This effectively means that CRs are uniformly coupled to the gas, and can supply momentum to all phases. This second case is abbreviated as the \nlX case. The third, fiducial case includes two-moment \crmhd with NLLD and IND. CRs decouple from cold neutral gas, which reduces their impact on the star formation rate and allows CRs to drive a warm steady-state galactic wind \sttfp. The third case is abbreviated as the \inX case. The \inX was previously designated as ``full-physics'' in \sttf. Due to the lack of explicit radiative transfer in this simulation, full-physics is a misnomer when analyzing these simulations in the context of star cluster formation. Instead, we refer to this case as the ``fiducial'' case for this work.

These simulations adopt the ``tallbox'' approach, with a domain of $1\;\mathrm{kpc}\times1\;\mathrm{kpc}\times\pm4\;\mathrm{kpc}$. The gas is initialized with a surface mass density of $10\;\mathrm{M}_\odot\;\mathrm{pc}^{-2}$ to be representative of solar neighborhood-like conditions. We set a target mass resolution of $10\;\mathrm{M}_\odot$ for the gas, ensuring that the star-forming ISM is resolved on parsec- and sub-parsec-scales.

For the first $\sim50\;\mathrm{Myr}$ of the simulation, we inject artificial SNe to prevent unphysical overcooling of the gas. Afterwards, the simulations tend toward a steady-state star formation rate mediated by supernova feedback. We analyze the simulations at $t\gtrsim112\;\mathrm{Myr}$, well after the imprint of this artificial SN-stirring phase has disappeared. These simulations run until $t\sim250\;\mathrm{Myr}$.

Within this analysis period, the three cases demonstrate significantly different trends in star formation. The \mhdX case has the highest SFR, and correspondingly the highest fraction of gas that is actively star-forming. The \nlX case experiences a significant reduction in SFR due to the strong coupling between CRs and gas in the ISM. The \inX case lies in-between these two cases; CRs with IND are able to modestly reduce the amount of star-forming gas \sttfp.


\subsection{\textsc{Crisp} Gas Physics}
\label{sec:crisp_model}

The \textsc{Crisp} model considers several cooling and heating processes that dominate the thermal evolution of cold gas. Specifically, we track diatomic, neutral, and ionized hydrogen. We include cooling from low-temperature fine-structure metal lines from ground state and singly ionized C, O, and Si, based on collision rates \citep{abrahamsson_fine-structure_2007, grassi_krome_2014}. Ly$\alpha$ cooling by neutral hydrogen \citep{cen_hydrodynamic_1992}, rotation-vibrational lines from $\mathrm{H}_2$ \citep{moseley_turbulent_2021} are the other significant contributors to cooling at low temperatures. Heating comes from FUV absorption \citep{bakes_photoelectric_1994} and CRs \citep{pfrommer_simulating_2017}. In the \mhdX case, CR ionization is applied uniformly using ionization rates from \cite{glover_star_2007}. In the \crmhd cases, CR ionization rates are rescaled based on the local CR energy density. We find in our \crmhd simulations that the ISM CR energy densities are fairly uniform and close to the solar neighborhood value \sttfp. Therefore, in practice, CR ionization rates are nearly the same between all three cases. The remaining features of the \textsc{Crisp} feedback model are described in \citecrisp, and are outlined in \citet{thomas_why_2025} and \sttf.


\subsection{Star Particles}
\label{sec:star_particle_methods}

With the \textsc{Crisp} feedback model, we use a Schmidt-type approach to form star particles \citep{schmidt_rate_1959, kravtsov_origin_2003}. Cells with gas density above the threshold density $n_\mathrm{H}=10^{3}\;\mathrm{cm}^{-3}$ are designated as ``star-forming''. This choice of density threshold is motivated by the limitations of our thermochemistry model and resolution. In the cores of molecular clouds where gas densities exceed $n_\mathrm{H}=10^{3}\;\mathrm{cm}^{-3}$, $\mathrm{CO}$ formation can become appreciable \citep{glover_approximations_2012, walch_silcc_2015}, contributing to the thermochemical state of the gas \citep[c.f.,][]{oka_interstellar_2006}. $\mathrm{CO}$ is not included in the \textsc{Crisp} model, implying that thermochemistry at densities $n_\mathrm{H}>10^{3}\;\mathrm{cm}^{-3}$ is not described by our model. Additionally, observed high-density star-forming clumps develop substructure at sizes below our resolution scale ($10\;\mathrm{M}_\odot$ or $\sim0.5\;\mathrm{pc}$). Both the spatial and thermochemical evolution of molecular clouds at densities $n_\mathrm{H}>10^{3}\;\mathrm{cm}^{-3}$ are unresolved by our model; therefore, we adopt the threshold density $n_\mathrm{H}=10^{3}\;\mathrm{cm}^{-3}$ to abstract these unresolved processes away into our simple subgrid star-formation prescription.

We use a local star formation efficiency per free-fall time of $100\%$. Gas cells with $n_\mathrm{H}\geq10^3\;\mathrm{cm}^{-3}$ are rapidly turned into star particles at a characteristic timescale corresponding to the freefall time $t_\mathrm{ff}\propto(G\rho)^{-1/2}$, so we ensure that gas does not reach extremely high densities beyond the scope of our model. Star formation is implemented stochastically, and the probability of a star-forming gas cell becoming a star particle is described in \citet{springel_cosmological_2003}. When a gas cell is set to become a star, the entire mass of the gas cell is instantaneously converted into a collisionless star particle. The massless gas cell is removed, and the nearby gas cells fill in the empty space at the next Voronoi mesh construction \citep{springel_e_2010}. The star particle inherits the mass of its progenitor gas cell, meaning that star particles in these simulations have masses of approximately $10\;\mathrm{M}_\odot$. This simple approach has been shown to produce reasonable star-formation properties in Lagrangian codes such as \textsc{Arepo} \citep{buck_observational_2019, keller_empirically_2022, hu_code_2023}.

The gravitational potential for star particles in \textsc{Arepo} is the collisionless particle potential defined in \citet{springel_e_2010} and \citet{weinberger_arepo_2020}. The gravitational softening length for star particles in our simulations is set to $1\;\mathrm{pc}$, meaning that the potential becomes Newtonian at a separation of $2.8\;\mathrm{pc}$. This, notably, is comparable to the typical radius of a star cluster \citep[several $\mathrm{pc}$; e.g.,][]{brown_radii_2021}, implying that the gravitational interactions between stars within an individual star cluster are not resolved. Because of this limitation, we do not analyze the precise internal dynamics of these star clusters, and instead focus on integrated properties. We discuss the minor impacts of the gravitational softening length on our results in Appendix~\ref{sec:app_discuss_limitations}.

\subsection{Stellar Feedback}
\label{sec:stellar_feedback_methods}

We include stellar feedback in the form of SNe. We do not explicitly track star particles through an initial mass function or mass-based evolutionary tracks. Instead, we aim to reproduce integrated quantities of stellar feedback. Any star particle older than $3\;\mathrm{Myr}$ and younger than $38\;\mathrm{Myr}$ has a probability to become a SNe. The fraction of star particles that become SNe is defined such that there is $1$ SNe per $100\;\mathrm{M}_\odot$ of stellar mass formed. Each SNe injects $1.06\;\times10^{51}\;\mathrm{erg}$ of energy. In the two \crmhd cases, an extra $5\%$ of the SN energy is included as injected CRs.

We do not include any direct radiative feedback from massive stars or (proto-)stellar winds. This is known to change the properties of early molecular cloud disruption and the clustering of SNe \citep[e.g.,][]{gutcke_lyra_2021, rathjen_silcc_2021, smith_efficient_2021, calura_siege_2025}. Regarding SN clustering, we focus instead on relative differences between our three cases. We expect that the effects of radiative feedback and CR feedback on SN clustering and molecular cloud disruption can be considered independently \citep[e.g.,][]{rathjen_silcc_2021, fitz_axen_suppressed_2024}. Therefore, the relative differences between our three cases are likely robust to the inclusion of radiative feedback. To verify this assumption, we provide order-of-magnitude estimations to infer the hierarchy of the various discussed feedback mechanisms.

Consider $t=0$ to be the time of formation of the first massive star in a star cluster. At $t=0$, this massive star (followed soon by other massive stars) begins releasing significant amounts of photoionizing radiation, creating photoioninizing ``fronts.'' The member stars of the newly-forming cluster, however, are still embedded within the dense progenitor molecular cloud. In dense environments, the propagating photoionization fronts surrounding massive stars generally only reach several parsecs in radius \citep{stromgren_physical_1939}---well within the scale of molecular clouds. Thus, during the early ``embedded'' stage of cluster formation, radiative feedback acts primarily locally on the progenitor molecular cloud. Stellar winds (also beginning during the embedded stage of cluster formation) are known to be energetically important, but likely less so than early radiative feedback \citep[e.g.,][]{rathjen_silcc_2021, kim_photochemistry_2023}. Additionally, the momentum contribution of stellar winds is quickly overshadowed by the onset of SNe.

When $t\geq3\;\mathrm{Myr}$, some sufficiently old stars become SNe, releasing significant energy and momentum into their surroundings. The mechanical energy from SNe can catastrophically disrupt nearby dense gas, particularly still-star-forming regions in the progenitor molecular cloud. In molecular gas of characteristic density $n_\mathrm{H}\sim100\;\mathrm{cm}^{-3}$, the energetic Sedov-Taylor blast wave stage will reach an approximate radius of $\lambda_\mathrm{SN}^*\sim2.85\;\mathrm{pc}$ \citep{blondin_transition_1998}. Subsequent SNe can explode within the bounds of previous supernova remnants, injecting energy into an already-diffuse environment and allowing for the energetic blast wave to propagate further, disrupting more gas. Star formation then becomes self-limiting. New stars can become SNe, and these SNe have increasingly significant feedback effects due to the preconditioning of the environment by previous SNe. Star cluster formation will therefore certainly cease several $\mathrm{Myr}$ after the onset of SNe, if star formation has not already been halted by local radiative feedback.

Under our assumptions, CRs are injected concurrently with SNe, and therefore cannot be injected until at least $t\geq3\;\mathrm{Myr}$. Because of this delay, CRs are not an immediate local feedback mechanism during the early embedded stage. CRs with $\sim\mathrm{GeV}$ energies can undergo several ``cooling'' processes resulting in net heating of the gas, namely hadronic, Coulomb, and ionization losses. These processes are essentially uniform under solar neighborhood conditions due to the extremely small effective cross sections $\sigma\lesssim10^{-25}\;\mathrm{cm}^{-2}$ \citep{padovani_cosmic-ray_2009, pfrommer_simulating_2017, ruszkowski_cosmic_2023}. The energetic contributions of CRs to the gas are therefore not local feedback, but rather global. The more significant impact of CR feedback is delivered through momentum transfer from CRs to gas. CRs exchange momentum with magnetized gas via the gyroresonant instability \citep{zweibel_basis_2017}, with scattering rates varying significantly depending on gas phase and environment \citep[e.g.,][]{armillotta_cosmic-ray_2021, thomas_effective_2025, hix_dynamically_2025}.  When ion-neutral damping is considered (as in the \inX case), gyroresonant interactions are significantly hindered in primarily neutral regions, resulting in little momentum transfer between CRs and gas. This can lead to mean-free-paths for CR momentum transfer of $\lambda_\mathrm{cr,p,IND}^\star\gtrsim100\;\mathrm{pc}$ in gas of $n_\mathrm{H}\sim100\;\mathrm{cm}^{-3}$ \citep{armillotta_cosmic-ray_2021}, or much greater \citep[e.g.,][]{thomas_effective_2025}. Therefore, in the \inX case, CRs certainly do not contribute to the disruption of star formation in embedded star clusters. Without ion-neutral damping (the \nlX case), gyroresonant interactions are frequent in dense gas \citep[limited only by nonlinear Landau damping; e.g.,][]{armillotta_cosmic-ray_2021, thomas_effective_2025}, giving momentum transfer mean-free-paths of $\lambda_\mathrm{cr,p}^\star< 1\;\mathrm{pc}$. However, despite the appearance of significant small-scale effects, the actual amount of momentum transferred from CRs to the gas still depends on the CR pressure gradient. Because the CR pressure profile will depend on gas motions and local magnetic topology (which themselves depend on the momentum contribution by CRs), CR feedback must be constrained through comprehensive modeling and sophisticated simulations.

Altogether, the hierarchy of scales is as follows: Radiation and stellar winds dominate for $t<3\;\mathrm{Myr}$, delivering energy to the immediate few parsecs surrounding the young star cluster. When $t\geq3\;\mathrm{Myr}$, some stars become SNe, which, depending on the density of the environment, can create remnants with sizes of a few parsec or larger. Subsequent SNe have greater feedback effects due to preconditioning by prior feedback, depositing energy further and further from the SN site. By a combination of stellar wind, radiation, and SN feedback, the parent molecular cloud is disrupted and star formation ceases at $t\sim10\;\mathrm{Myr}$. Alongside SNe, CRs are accelerated. When ion-neutral damping is considered, these CRs rapidly escape their acceleration site without delivering much energy or momentum. In the absence of ion-neutral damping, these CRs \textit{may} contribute a small amount of energy and momentum to the star-forming cloud. Regardless, CR feedback likely does not contribute to local disruption of star formation. Instead, CRs pollute the ISM, develop gradients over scales larger than clouds \citep{kjellgren_dynamical_2025}, and deliver momentum to the galactic wind \sttfp, mostly independent from the local effects of radiation and stellar winds \citep{rathjen_silcc_2021}. Because our simulations omit direct radiative feedback and stellar winds, we neglect local feedback for the first $3\;\mathrm{Myr}$ of cluster formation; however, we still capture the ultimate disruption of cluster formation by successive SNe and the global effects of CRs on the ISM.


\section{Star Particle Distribution}
\label{sec:star_particles_with_pretty_figure}

\begin{figure*}
\includegraphics{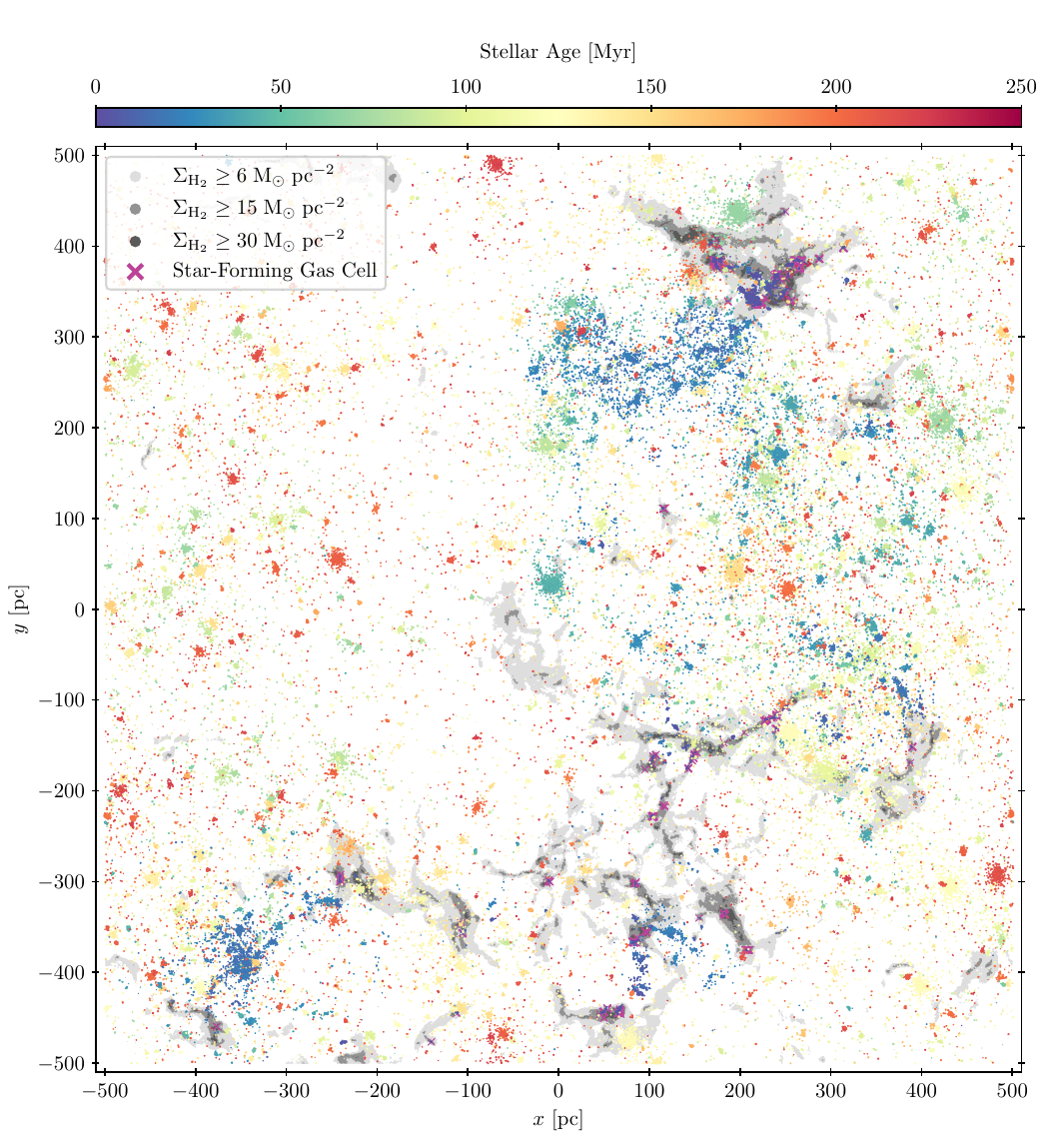}
\caption{Star particles in the \inX simulation at $t\approx250\;\mathrm{Myr}$, colored by the age of the star particle. Bluer particles are younger, redder particles are older. Three filled contours of molecular hydrogen surface density are displayed in gray, with star-forming gas cells marked with a purple $\times$. We see a variety of clusters of star particles with similar ages, suggesting these stars formed together and remained bound. The youngest star clusters are primarily localized near molecular clouds and star-forming regions. In contrast, the oldest stars and star clusters appear throughout the full domain.
\label{fig:pretty_star_plot}}
\end{figure*}

The face-on distribution of star particles and molecular column density contours in the \inX simulation at $t=250.31\;\mathrm{Myr}$ are shown in Figure~\ref{fig:pretty_star_plot}. We color the star particles by their age to emphasize any trends between location and stellar age.

We immediately see clusters of similarly-aged particles spread throughout the domain. Clusters of many different ages are visible, including a $\sim200\;\mathrm{Myr}$-old star cluster at $x\approx-250\;\mathrm{pc},y\approx50\;\mathrm{pc}$, a $\sim60\;\mathrm{Myr}$-old cluster at $x\approx-5\;\mathrm{pc},y\approx5\;\mathrm{pc}$, and a still-forming cluster embedded within a molecular cloud at $x\approx250\;\mathrm{pc},y\approx350\;\mathrm{pc}$.

These clusters of star particles are spatially compact and consist of similarly-aged particles. Additionally, some fairly old clusters exist with ages over $100\;\mathrm{Myr}$, suggesting that these clusters can remain gravitationally bound in this simulation for longer than $100\;\mathrm{Myr}$. This matches the definition for star clusters; therefore, we designate these simulated associations of star particles to be ``star clusters.''

The oldest star clusters (``redder'' clusters in Figure~\ref{fig:pretty_star_plot}) exist throughout the full domain. These clusters have had ample time to move throughout the simulation box and spatially de-correlate from the star-forming regions in which they were born. In contrast, the youngest star clusters (with ages $\lesssim20\;\mathrm{Myr}$; ``bluer'' clusters in Figure~\ref{fig:pretty_star_plot}) exist only within a handful of regions, particularly near molecular clouds. These clusters have likely recently dispersed their progenitor GMCs via SN feedback, but have not yet had time to disperse throughout the domain.

Star-formation occurs primarily within the highest surface density ($\Sigma_{\mathrm{H}_2} \geq30\;\mathrm{M}_\odot\;\mathrm{pc}^{-2}$) cores of molecular clouds. This is interesting to note, when we recall that our star formation criteria is simply a density threshold ($n_\mathrm{H}>10^{3}\;\mathrm{cm}^{-3}$), and contains no extra criteria regarding molecular hydrogen. The correlation between star-forming regions and regions of molecular hydrogen is not enforced, but rather a satisfying result from this simulation. As clouds become sufficiently dense, molecular hydrogen begins to form and gravitational forces dominate, and star formation follows rapidly \citep{hartmann_rapid_2001}. This points to the robustness of the \textsc{Crisp} feedback model, despite the simple star-formation prescription. Star particles form within the dense interiors of molecular clouds \citep{krumholz_star_2019}, resulting in localized star clusters that form over a single short epoch. These star clusters quickly disperse their surroundings, and can remain bound for over $100\;\mathrm{Myr}$. No star formation occurs outside of the dense clouds, owing to our high star formation density threshold. Through the interplay of these various physical processes in \textsc{Crisp} alongside our simple star formation approach, our model is able to reproduce the formation of stars in apparently bound clusters. This result is distinct from subgrid star cluster formation models \citep[e.g.,][]{li_star_2017, reina-campos_star_2025}, where the formation of bound star clusters is hard-coded into the model.
We do not prescribe clustered star formation; rather, we identify star clusters as a result of the model. The remaining analysis in this work tests the extent to which our simulated star clusters are representative of reality.

The primary purpose of Figure~\ref{fig:pretty_star_plot} is to show that star clusters exist in these simulations, and to motivate the rest of this work. We do not include analogous face-on distributions of star particles and molecular column density from the \mhdX and \nlX cases, because these additional figures would be superfluous in motivating the remaining analysis.


\section{Mass Functions}
\label{sec:mass_functions_section}

For this section, we focus on the mass functions of young star clusters and the environmental dependence. We analyze the mass functions for our three simulations, and we compare with observational data.


\subsection{Cluster Identification}
\label{sec:cluster_identification}

We identify star clusters through a modified Friends-of-Friends (FoF) algorithm. We consider the formation time of star particle $i$ as a fourth coordinate $t_{*,i}$. Combining this time coordinate with the ordinary spatial coordinates $x_i$, $y_i$, and $z_i$, we define the vector $\bm{w}_i$ as
\begin{equation}
    \bm{w}_i\equiv\left(x_i, y_i, z_i, u_0\, t_{*,i}\right),
\end{equation}
for each star particle at a given snapshot, where $u_0 \equiv 1\, \mathrm{pc}/ 2\,\mathrm{Myr}$ is the conversion factor from time coordinates to spatial coordinates. The separation between two coordinates $\bm{w}_1$ and $\bm{w}_2$ is simply the Euclidean norm of their difference, treating all four dimensions equally. Our value of $u_0$ implies that a difference in age of $2\;\mathrm{Myr}$ is equivalent to a spatial separation of $1\;\mathrm{pc}$. By including this time coordinate in $\bm{w}_i$, we ensure that star clusters must consist of similarly-aged stars, and chance interlopers are not included. The selection of this value and its impact are discussed in Appendix~\ref{sec:app_discuss_fof}.

For the FoF algorithm, we use a fixed linking length $\Delta w=2\;\mathrm{pc}$. The selection of this parameter is discussed in Appendix~\ref{sec:app_discuss_fof}. Following the definition of $\bm{w}_i$, two coeval star particles separated by $2\;\mathrm{pc}$ will be joined in an FoF group, and two co-spatial star particles with an age difference of $4\;\mathrm{Myr}$ will also be joined in an FoF group. The FoF cluster identification proceeds in the normal fashion with this fixed linking length. After the FoF catalog for each snapshot is complete, we remove groups with less than 10 members, as they would represent loose associations of less than $100\;\mathrm{M}_\odot$.

We build a ``young cluster sample'' for each simulation case by parsing through the FoF catalogs from each snapshot, and selecting star clusters $>10\;\mathrm{Myr}$ after $t_{60}$, the time when the cluster only contained $60\;\mathrm{M}_\odot$ of stellar mass. This condition is implemented to ensure that early outlier star formation events do not dictate the age of a star cluster. We find that the oldest few star particles in a star cluster were often formed several $\mathrm{Myr}$ before the bulk star formation epoch. The time $t_{60}$ is used to ensure that the age of clusters is determined relative to the onset of the primary star formation period, rather than the formation of the first few outliers. The population-level properties of the young cluster sample are not sensitive to the adopted mass threshold of $60\;\mathrm{M}_\odot$ (and subsequent time determination $t_{60}$). By selecting clusters $\gtrsim 10\;\mathrm{Myr}$ after $t_{60}$, we sample them as their star-forming epoch is effectively completed (see Appendix~\ref{sec:app_formation_epoch} for further discussion).

We ensure that each young star cluster is only included in the sample once by masking out member star particles which are already included in the sample, constituting a ``uniqueness'' criterion. Candidate young clusters may not contain more than 10 star particles that have already been included in the sample. Therefore, a cluster is only included in the sample once, at the first snapshot that it satisfies the age criterion. The precise value of this threshold does not have a significant impact on our results. We select the value of 10 for our threshold to represent our effective resolution limit.

In addition to the uniqueness and minimum age criteria, we apply a maximum age criterion of $20\;\mathrm{Myr}$ (or $t-t_{60}\leq20\;\mathrm{Myr}$). However, because clusters are identified at the first possible snapshot, no valid clusters remain to be identified at this age. Thus, the maximum age criterion is only ever relevant at the first analysis snapshot ($t\approx112\;\mathrm{Myr}$). In this first snapshot, all valid clusters with $10\;\mathrm{Myr}<t-t_{60}\leq20\;\mathrm{Myr}$ are identified in the sample. In all following snapshots, the uniqueness criterion can be applied.


\subsection{Cluster Mass Functions}
\label{sec:mass_function_sub}

\begin{figure*}
\includegraphics{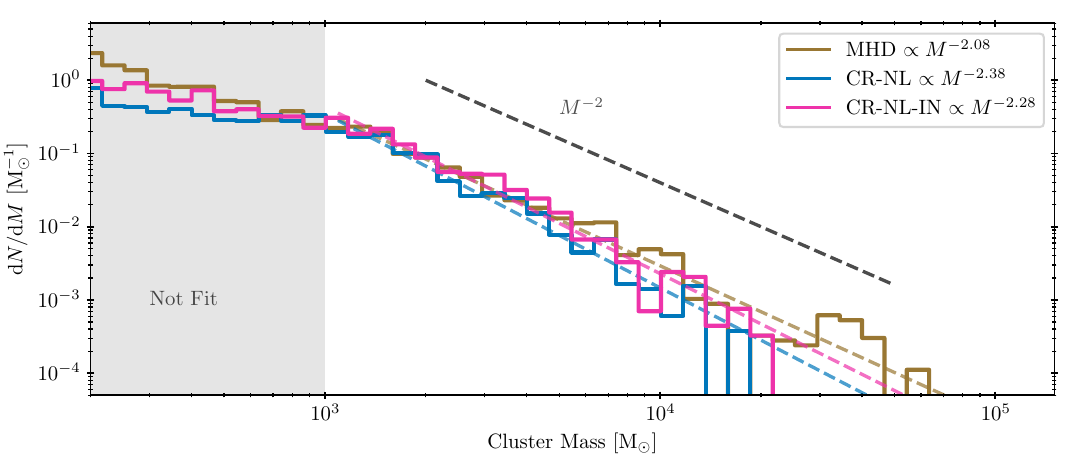}
\caption{Star cluster mass function using the young cluster sample described in Section~\ref{sec:cluster_identification}, where the star clusters are identified with an age of $\sim10\;\mathrm{Myr}$. The mass functions are fit with power law using a Poisson likelihood, as described in Section~\ref{sec:mass_function_sub}. The best fits are displayed as a dashed line, and the power law slope for each fit is reported in the legend. We neglect star clusters below a mass of $10^{3}\;\mathrm{M}_\odot$ in the fit, as the mass function appears to change slope at masses less than this threshold. We also include a reference line with power law slope $-2$. Each run produces a cluster mass function with a power law slope between $-2$ and $-2.5$, consistent with observationally-determined slopes in environments with $\Sigma_\mathrm{SFR}$ values similar to this work \citep[e.g.,][]{adamo_probing_2015, johnson_panchromatic_2017, adamo_star_2020, messa_young_2018}. Our results are reported in Table~\ref{tab:data_table_appendix}.
\vspace{3mm}
\label{fig:mass_function}}
\end{figure*}

The observed mass function for young star clusters in the nearby galaxies follows a power-law distribution with a slope of approximately $-2$, depending on the SFRD \citep{adamo_star_2020}. We fit the mass function $\mathrm{d} N / \mathrm{d} M$ of our young cluster samples above a mass of $10^{3}\;\mathrm{M}_\odot$ as a power law of the form
\begin{equation}
    \frac{\mathrm{d} N}{\mathrm{d} M} = B_0 \, M^{\alpha_\mathrm{PL}},
\end{equation}
where $\alpha_\mathrm{PL}$ is the slope and $B_0$ is the normalization. We use a Poisson likelihood for the fit and estimate errors via bootstrapping.

We do not include a more complex form of the mass function, such as a Schechter-like function with an exponential cutoff above a certain mass $M_\mathrm{c}$, because our simulations produce relatively low-mass clusters that fall below the typical values $M_\mathrm{c} \sim 10^{6-7}\; \mathrm{M}_\odot$ \citep[e.g.,][]{johnson_panchromatic_2017}. Additionally, the exponential cutoff strongly impacts the effective slope of the mass function within 1~dex of the cutoff mass, meaning $M_\mathrm{c}$ and $\alpha_\mathrm{PL}$ cannot be simultaneously fit with data that only span approximately 1~dex \citep{johnson_panchromatic_2017}.

The mass functions in the three cases, as well as power-law fits above $10^{3}\;\mathrm{M}_\odot$, are displayed in Figure~\ref{fig:mass_function}. We also include a reference power law $M^{-2}$. We do not fit the mass function below $10^{3}\;\mathrm{M}_\odot$ due to the clear spectral break at this value. Additionally, clusters below this mass may be especially sensitive to numerical effects, so we do not analyze them further. They are also typically not included in the observational samples. The power law slopes (with bootstrapped errors) are reported in Appendix~\ref{sec:app_data_table}, Table~\ref{tab:data_table_appendix}.

Firstly, we see that each case's cluster mass function is consistent with a single power law in the $10^{3}\;\mathrm{M}_\odot$ to $\lesssim10^{5}\;\mathrm{M}_\odot$ mass range. These power laws are all slightly steeper than the reference $M^{-2}$ slope. This result alone is remarkable. Both the manifestation of a power law star cluster mass function and the exact resulting slope are very sensitive to the feedback prescription \citep[e.g.,][]{andersson_pre-supernova_2024}. We produce plausible power law mass functions within our idealized tallbox setup, and without fine-tuning of feedback parameters. It is remarkable that the \textsc{Crisp} model recovers the power law form of the star cluster mass function, especially considering that we omit pre-SN feedback entirely. Recent work of \citet{andersson_pre-supernova_2024} has suggested that the manifestation of the power law form of the cluster mass function requires pre-SN feedback via radiation and stellar winds. It is possible that we emulate this result with our rapid prescription for star formation and SN feedback, as we utilize a star formation efficiency per freefall time of $100\%$, and a minimum stellar lifetime (age of a star particle that is allowed to become a SN) of $3\;\mathrm{Myr}$. However, \citet{smith_efficient_2021} found that pre-SN photoionizing radiative feedback can disrupt star-forming clouds before large clusters are able to form. Radiative feedback also reduces star cluster masses in \citet{rathjen_silcc_2021}, presumably by the same mechanism. It is possible that the massive clusters in our simulations are only able to form in the absence of radiative feedback, and that the power law form of the mass functions in our simulations is circumstantial. We emphasize that the value of this work is in comparisons between the three cases---understanding how CRs affect the formation of star clusters. As described in Section~\ref{sec:stellar_feedback_methods}, radiative feedback disrupts star cluster formation on local scales, whereas CRs tend to act on ISM-scales far away from sources. Including radiative feedback would likely change the three cases uniformly, leaving relative comparisons unaffected.

Looking at the high-mass end of the mass functions in Figure~\ref{fig:mass_function}, we see that the \mhdX case produces the highest mass star clusters, followed by the \inX case and then the \nlX case. This mirrors the SFRs from \sttf (the \mhdX case has the highest SFR, followed by the \inX case and the \nlX case). In our simulations, a higher SFRD is correlated with a larger most massive star cluster, consistent with other simulations \citep[e.g.,][]{li_star_2018} and observations \citep[e.g.,][]{adamo_probing_2015, johnson_panchromatic_2017}.


\subsection{Environmental Dependence}
\label{sec:environmental_dependence}

The star cluster mass function is known to vary with environmental properties, such as the local SFRD. To investigate this dependence, we split the young cluster sample into three subsets. 

First, we analyze the SFRD for each simulation as a function of time, and split it into low, medium, and high percentiles. This means that, by construction, the SFRD in each simulation was low, medium, and high for exactly $1/3$ of the analysis time. In this way, because observations occur at a single point in time, each of these bins would be equally likely to be ``observed.'' Once these SFRD bins are defined, we look at the SFRD at the time of the median stellar formation time of each cluster, within a window of $\pm0.5\;\mathrm{Myr}$. Based on the average SFRD within this window, each star cluster is placed into the low, medium, or high SFRD subsets.

\begin{figure}
\includegraphics{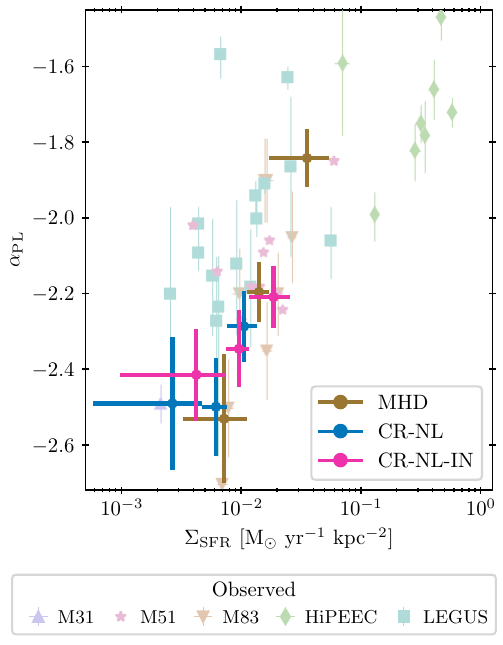}
\caption{The best fit and error for the power law slope of the mass function within an SFRD bin for the three cases, from this work and relevant observational works. The errors here are determined by bootstrapping. The power law slopes inferred from the simulations show a positive correlation with the SFRD and are broadly consistent with the observational values and trends. Observed values are from M31 \citep{johnson_panchromatic_2017}, M51 \citep{messa_young_2018}, M83 \citep{adamo_probing_2015}, the HiPEEC sample \citep{adamo_star_2020-1}, and the LEGUS sample \citep{adamo_star_2020}. Our results are reported in Table~\ref{tab:data_table_appendix}.
\label{fig:alpha_vs_sfrd}}
\end{figure}


\subsubsection{Power Law Slope}
\label{sec:power_law_slope_subsub}

The first environmentally-dependent aspect of star cluster formation we study is the power law slope $\alpha_\mathrm{PL}$ of the star cluster mass function. We calculate the power law slope within each of these three SFRD subsets and display the results in Figure~\ref{fig:alpha_vs_sfrd}. The numbers are also reported in Table~\ref{tab:data_table_appendix}.

Looking first at our data as a whole, we see a clear positive correlation between $\alpha_\mathrm{PL}$ and $\Sigma_\mathrm{SFR}$, with moderate scatter. This is consistent with observations \citep[e.g.,][]{adamo_probing_2011, adamo_star_2020}. Next, we look at the trends for $\alpha$ within the individual cases. The data points from the \mhdX case follow a fairly steep, linear relation in $\alpha_\mathrm{PL}-\log \Sigma_\mathrm{SFR}$. The data points from the \inX case follow a shallower, but still positive, trend. The \nlX case is unique in that its lowest SFRD bin does not have the lowest value for $\alpha_\mathrm{PL}$. This can be attributed to the intrinsic scatter and uncertainty on $\alpha_\mathrm{PL}$. Indeed, the errors determined from bootstrapping permit a positive trend.

Our inferred values for $\alpha_\mathrm{PL}$ vs. $\Sigma_\mathrm{SFR}$ are consistent with the observational points in Figure~\ref{fig:alpha_vs_sfrd}. Although our data seem to be on the lower end of $\alpha_\mathrm{PL}$ for a given $\Sigma_\mathrm{SFR}$, relative to the observational data, it is still very consistent with the observations of M83 \citep{adamo_probing_2015}, M31 \citep{johnson_panchromatic_2017}, and M51 \citep{messa_young_2018}. Interestingly, adding CRs (with or without ion-neutral damping) simply seems to move the data points within the observationally-permitted range, rather than producing unphysical results. Therefore, we cannot use the results in Figure~\ref{fig:alpha_vs_sfrd} to determine which simulation is the best match to reality.


\subsubsection{Cluster Formation Efficiency}
\label{sec:cfe_subsub}

The fraction of stellar mass formed in bound star clusters (relative to the total stellar mass formed) is known as the CFE (or $\Gamma$), and is widely found to depend on the environment. We calculate the CFE using a four-step process. First, we assign star clusters to a SFRD bin. Next, we calculate the stellar mass that is gravitationally bound within each cluster. Third, we apply a minimum mass cut (either $1000\;\mathrm{M}_\odot$ or $5000\;\mathrm{M}_\odot$) to account for resolution and to emulate other work \citep{goddard_fraction_2010, adamo_probing_2015, li_star_2017}. Finally, we calculate the total bound cluster mass formed within a SFRD bin, and divide it by the total stellar mass formed within the SFRD bin to get the CFE. We elaborate on the individual steps below.

We use an iterative approach to identify clusters that are bound by their self-gravity. First, we isolate the FoF-identified members of the cluster. Next, we calculate the gravitational potential energy between every pair of particles. For this, we use the softening kernel from the simulation, to ensure that the gravitational binding calculation is self-consistent. We then compare each particle's potential energy with its kinetic energy (relative to the center of mass of the cluster). Particles with kinetic energy exceeding their potential energy are considered unbound, and removed from the cluster. This gravitational binding calculation is repeated with members that were determined to be bound, until a converged subset of star particles is identified. These star particles are then considered to be ``converged members'' of the gravitationally-bound star cluster. This approach only considers the self-gravity of the star cluster, and does not take into account tidal interactions with nearby GMCs \citep[the ``cruel cradle effect,''][]{kruijssen_fraction_2012}, or evaporation over the lifetime of the cluster.

\begin{figure}
\includegraphics{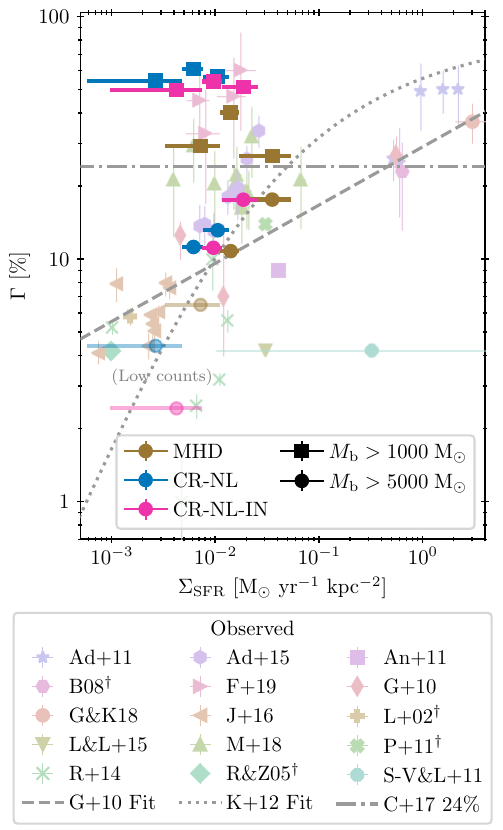}
\caption{CFE ($\Gamma$) vs. SFRD range ($\Sigma_\mathrm{SFR}$) for the three cases and three SFRD subsamples. The CFE is defined as the percent of stellar mass that is formed in massive bound clusters ($M_\mathrm{bound}>1000\;\mathrm{M}_\odot$, squares; $M_\mathrm{bound}>5000\;\mathrm{M}_\odot$, circles) relative to the total stellar mass formed in the SFRD bin. The bound cluster criteria is described in Section~\ref{sec:cfe_subsub}. The values of $\Gamma$ from our simulations with the lower mass threshold of $5000\;\mathrm{M}_\odot$ are an excellent match to the theoretical prediction from \citet{kruijssen_fraction_2012} and broadly agree with the data from observations. Observed values, in order of appearance in the legend, are from \citet{adamo_probing_2011}, \citet{adamo_probing_2015}, \citet{annibali_cluster_2011}, \citet{bastian_star_2008}, \citet{fensch_massive_2019}, \citet{goddard_fraction_2010}, \citet{ginsburg_high_2018}, \citet{johnson_panchromatic_2016}, \citet{larsen_luminosity_2002}, \citet{lim_star_2015}, \citet{messa_young_2018}, \citet{pasquali_infrared_2011}, \citet{ryon_snapshot_2014}, \citet{rafelski_star_2005}, and \citet{silva-villa_star_2011}. Fits are from \citet{goddard_fraction_2010}, \citet{kruijssen_fraction_2012}, and \citet{chandar_fraction_2017}. Labels with a $\dagger$ symbol in the legend have SFRD values determined from their respective references and $\Gamma$ values from \citet{goddard_fraction_2010}. Our results are reported in Table~\ref{tab:data_table_appendix}.
\label{fig:gamma_vs_sfrd}}
\end{figure}

To determine the CFE, we first calculate the bound mass for each cluster. We perform this analysis twice, with different minimum mass thresholds for bound clusters--$1000\;\mathrm{M}_\odot$ and $5000\;\mathrm{M}_\odot$. We use $5000\;\mathrm{M}_\odot$ as a comparison for the lower completeness limit of many observations \citep[e.g.,][]{adamo_probing_2015, messa_young_2018}, and we use $1000\;\mathrm{M}_\odot$ to demonstrate the dependence of our results on the selected lower mass threshold. Next, we calculate the total stellar mass formed within each SFRD bin (e.g., $M_\star = \int \mathrm{SFR}\;\mathrm{d}t=\langle\mathrm{SFR}\rangle \Delta t$). The CFE for each SFRD bin is simply the total mass formed in bound clusters with masses greater than the threshold, divided by the total stellar mass formed in that SFRD bin. The CFEs calculated with minimum masses $1000\;\mathrm{M}_\odot$ and $5000\;\mathrm{M}_\odot$ are reported as percentages $\Gamma$ in Figure~\ref{fig:gamma_vs_sfrd}, and listed in Table~\ref{tab:data_table_appendix}.

We first focus on the $\Gamma$ values calculated using a minimum mass of $5000\;\mathrm{M}_\odot$ (circles in Figure~\ref{fig:gamma_vs_sfrd}). The simulated data as a whole show a positive trend between $\Gamma$ and $\Sigma_\mathrm{SFR}$. Each case also shows the same trend individually between the three SFRD bins. This means that a larger SFRD is correlated with a greater fraction of stellar mass formed in bound, massive ($M_\mathrm{cl}>5000\;\mathrm{M}_\odot$) star clusters. Again similar to Figure~\ref{fig:alpha_vs_sfrd}, we see in Figure~\ref{fig:gamma_vs_sfrd} that adding CRs only moves the resulting data points within the range of observations, and along the predicted relation. Figure~\ref{fig:gamma_vs_sfrd}, therefore, may offer no constraining power in determining which simulation is the most accurate.

Our results are in fair agreement with many points from observations \citep{adamo_probing_2015, messa_young_2018}, and in good agreement with the theoretical prediction from \citet{kruijssen_fraction_2012}. This is a somewhat unexpected result, as we do not capture small-scale gravitational interactions that could alter boundedness. Additionally, our methodology is very different from most observational works. Generally, in observational works, the low-mass end of the cluster mass function is assumed to be incomplete, and the contribution of cluster mass from undetected low-mass clusters is calculated by assuming a power law form for the cluster mass function and integrating it to a low ($\sim100\;\mathrm{M}_\odot$) threshold \citep[e.g.,][]{adamo_probing_2015, messa_young_2018}. We instead simply reject clusters below $5000\;\mathrm{M}_\odot$ and sum the mass of the remaining clusters. To determine whether our findings are meaningful, we must investigate further.

When we instead use the lower mass threshold of $1000\;\mathrm{M}_\odot$, we see much different results. The $\Gamma$ values do not follow a positive trend with $\Sigma_\mathrm{SFR}$. Rather, the two \crmhd cases are consistent with a high, constant $\Gamma$. The \mhdX case has lower $\Gamma$ values than both \crmhd cases, and also displays a higher $\Gamma$ value for its intermediate $\Sigma_\mathrm{SFR}$. These $\Gamma$ values are all much higher than the majority of observations. 
The apparently high $\Gamma$ values inferred from our data can be partially explained by a combination of the relatively high star formation density threshold and $\epsilon_\mathrm{ff}$ used for our star formation prescription. With these parameters, stars form in localized dense gas \citep{buck_observational_2019, keller_empirically_2022}, reducing the possibility of stars forming outside of clusters. If these simulated $\Gamma$ values are indeed too large, they could be remedied by simultaneously decreasing the star formation density and $\epsilon_\mathrm{ff}$, allowing some stars to form in lower-density gas without significantly changing the star formation rates \citep[e.g.,][]{smith_efficient_2021, hislop_challenge_2022}. Direct radiative feedback may also act to decrease $\Gamma$ \citep{hislop_challenge_2022}. It is unclear which of our $\Gamma$ values is most correct or most applicable for comparison with observations due to non-uniform conventions in the literature \citep{chandar_fraction_2017}.

It is possible that the positive correlation between $\Gamma$ (using the threshold of $5000\;\mathrm{M}_\odot$), and the SFRD is purely artificial. If we assume that the bound cluster mass function is a power law that varies with the SFRD like the full (including unbound) young cluster mass function, the fraction of cluster mass above $5000\;\mathrm{M}_\odot$ will always be higher for a larger (less negative) $\alpha_\mathrm{PL}$, and therefore higher for a larger $\Sigma_\mathrm{SFR}$. Therefore, with increasing $\Sigma_\mathrm{SFR}$, we include a larger fraction of cluster formation when calculating the cluster formation rate, and therefore infer a higher CFE. Put simply, with increasing $\Sigma_\mathrm{SFR}$, the mass function is higher and extends further beyond $5000\;\mathrm{M}_\odot$. This effect happens in addition to the bound cluster formation rate of star clusters above $1000\;\mathrm{M}_\odot$. The bound cluster formation rates (equivalently the $\Gamma$ values) of clusters more massive than $1000\;\mathrm{M}_\odot$ do not seem to vary strongly with $\Sigma_\mathrm{SFR}$ within a single case, suggesting that the artificial variation due to our lower mass threshold of $5000\;\mathrm{M}_\odot$ could explain our results. More detailed analysis would be required to verify whether our $\Gamma-\Sigma_\mathrm{SFR}$ relation is physical or artificial.


\section{Clustering of Supernovae}
\label{sec:sn_clustering}

\begin{figure*}
\includegraphics{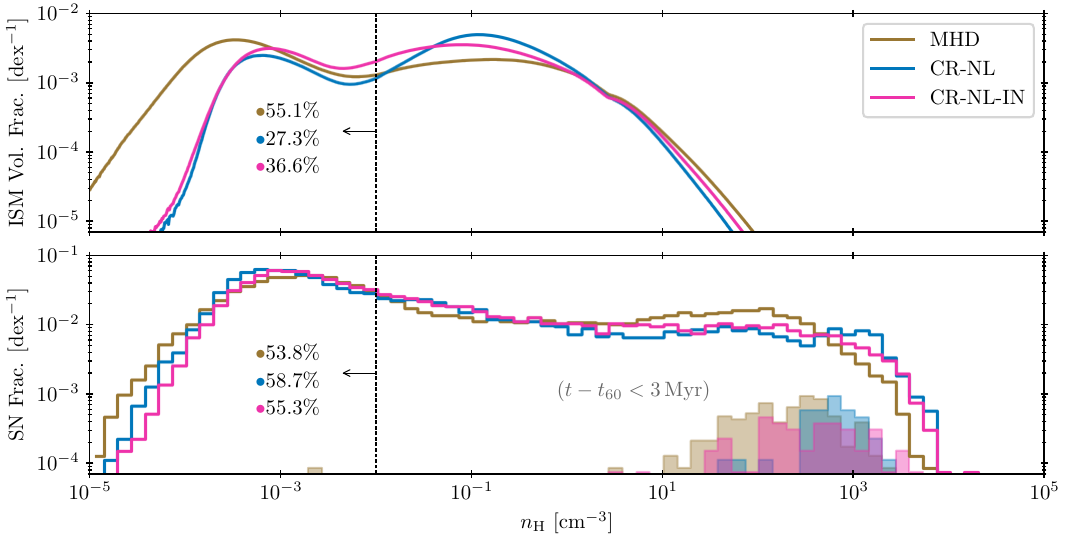}
\caption{\textit{Upper panel:} Distribution of ISM volume as a function of density ($n_\mathrm{H}$). The sample includes gas within $200\;\mathrm{pc}$ of the midplane over the time range $112\;\mathrm{Myr}$ to $250\;\mathrm{Myr}$. \textit{Lower panel}: Distribution of SN injection environment densities, over the same time range. The lightly shaded regions are the fraction that occur less than $3\;\mathrm{Myr}$ after $t_{60}$ (the approximate beginning of a cluster's formation). For both distributions, the fraction occurring below densities $10^{-2}\;\mathrm{cm}^{-3}$ are included as a percentage. Despite the significant differences in ISM gas distributions, approximately half of SNe occur in diffuse gas environments, irrespective of the CR model. Additionally, the first few SNe tend to occur in very dense gas, tracing the conditions of star particle formation. Pre-SN feedback (direct radiative feedback and stellar winds) would likely spread these first few SNe among lower-density gas \citep{rathjen_silcc_2021}.
\label{fig:sn_gas_density_tracing}}
\end{figure*}

When massive stars form in clusters, consecutive SNe are expected to occur within the diffuse, ``pre-conditioned'' ambient medium created by previous SNe. This preconditioning reduces the radiative losses experienced by an expanding supernova remnant, allowing a greater fraction of the initial energy to reach the ISM and create galactic winds. Ultimately, the effectiveness of a SN in delivering energy to the ISM is not directly described by the clustering of SNe, but instead the density of the ambient medium in which the SN occurs \citep{mccray_supershells_1987, gatto_modelling_2015}. In addition to SNe, radiative feedback from massive stars can also create low-density regions. These therefore act as an additional source of preconditioning. Our simulations lack direct radiative feedback; therefore, our conclusions relating to SN preconditioning are inherently limited. The degree of SN clustering is also reflective of the degree of stellar clustering \citep[e.g.,][]{smith_efficient_2021}, meaning that our choice of star formation density threshold and $\epsilon_\mathrm{ff}$ could play a role in determining the exact circumstances of SN clustering. However, CRs have larger-scale effects on the ISM (see Section~\ref{sec:stellar_feedback_methods} and \sttf), possibly altering the gas density distribution around star-forming sites. CR feedback could feasibly amplify or inhibit the effects of preconditioning independently from local-scale effects.

As in other works that present simulations with individual massive stars \citep[e.g.,][]{gutcke_lyra_2021, smith_efficient_2021}, we explore the distribution of ambient gas densities in which SNe occur. We show the distribution of ambient gas densities at SN injection sites for the three cases in Figure~\ref{fig:sn_gas_density_tracing} (bottom panel). For comparison, we also show the underlying volume-weighted density distribution of the ISM in Figure~\ref{fig:sn_gas_density_tracing} (top panel).

First, looking at the bottom panel of Figure~\ref{fig:sn_gas_density_tracing}, we see that the distribution of ambient gas densities at SN injection sites are remarkably similar between the three cases. There is a modest peak at low ($n_\mathrm{H}\leq10^{-2}\;\mathrm{cm}^{-3}$) densities (corresponding to the diffuse warm/hot phase of the ISM), and a large tail that extends just beyond the star-formation density threshold. The shapes of these distributions are consistent with simulations that do not include pre-SN radiative feedback \citep[e.g.,][]{gutcke_lyra_2021, rathjen_silcc_2021, smith_sensitivity_2021}. 

To quantify the fraction of SNe that can be considered ``clustered,'' we evaluate the fraction of SNe that occur in significantly diffuse ($n_\mathrm{H}\leq10^{-2}\;\mathrm{cm}^{-3}$) gas. This density approximately separates the diffuse warm/hot phase of the ISM from the colder, denser phases of the ISM (see Figure~\ref{fig:sn_gas_density_tracing}, top panel), so we approximately designate gas with densities below $10^{-2}\;\mathrm{cm}^{-3}$ to be pre-conditioned. The fractions of SNe that occur in pre-conditioned gas are reported as percentages inset in Figure~\ref{fig:sn_gas_density_tracing}. We see, quite remarkably, that approximately half of SNe occur in significantly diffuse gas in all three physics cases. Although there are slight differences between the three percentages, we do not consider the differences to be large enough to be scrutinized.

Next, we compare the ambient gas densities at SN injection sites to the underlying volume distribution of the ISM. If SNe were distributed randomly throughout the ISM, the ambient gas densities at SN injection sites (Figure~\ref{fig:sn_gas_density_tracing}, bottom panel) would be identical to the ISM volume-weighted density distribution (Figure~\ref{fig:sn_gas_density_tracing}, top panel). However, we see that this is certainly not the case. This implies that SNe in our simulations are not an unbiased tracer of the ISM. Rather, approximately half of the SNe in each simulation occur in diffuse ambient gas, independent of how the overall ISM has been shaped by feedback.

In Figure~\ref{fig:sn_gas_density_tracing} (bottom panel), we also show the distribution of ambient gas densities at SN injection sites for SNe that occur at the earliest stages of star cluster formation ($t-t_{60}<3\;\mathrm{Myr}$, where $t_{60}$ was defined in Section~\ref{sec:cluster_identification}) as a shaded histogram. In all three cases, this distribution is peaked around the star-forming density threshold $n_\mathrm{H}=10^{3}\;\mathrm{cm}^{-3}$, with some scatter owing to quasi-random wandering motions of young star particles. Because we include no radiative feedback, the first few SNe in a cluster will occur in the high-density, star-forming core of a molecular cloud \citep[e.g.,][]{rathjen_silcc_2021}. This is likely not reflective of reality, where radiative feedback and stellar winds result in some degree of early gas excavation. As time passes and more SNe occur, the ambient gas around a star cluster becomes sufficiently ``pre-conditioned'' (low-density), allowing SNe to populate the low-density region of the distribution in Figure~\ref{fig:sn_gas_density_tracing} \citep[e.g.,][]{gutcke_lyra_2021}.


\section{Star Cluster Properties}
\label{sec:cluster_properties}

\begin{figure*}
\includegraphics{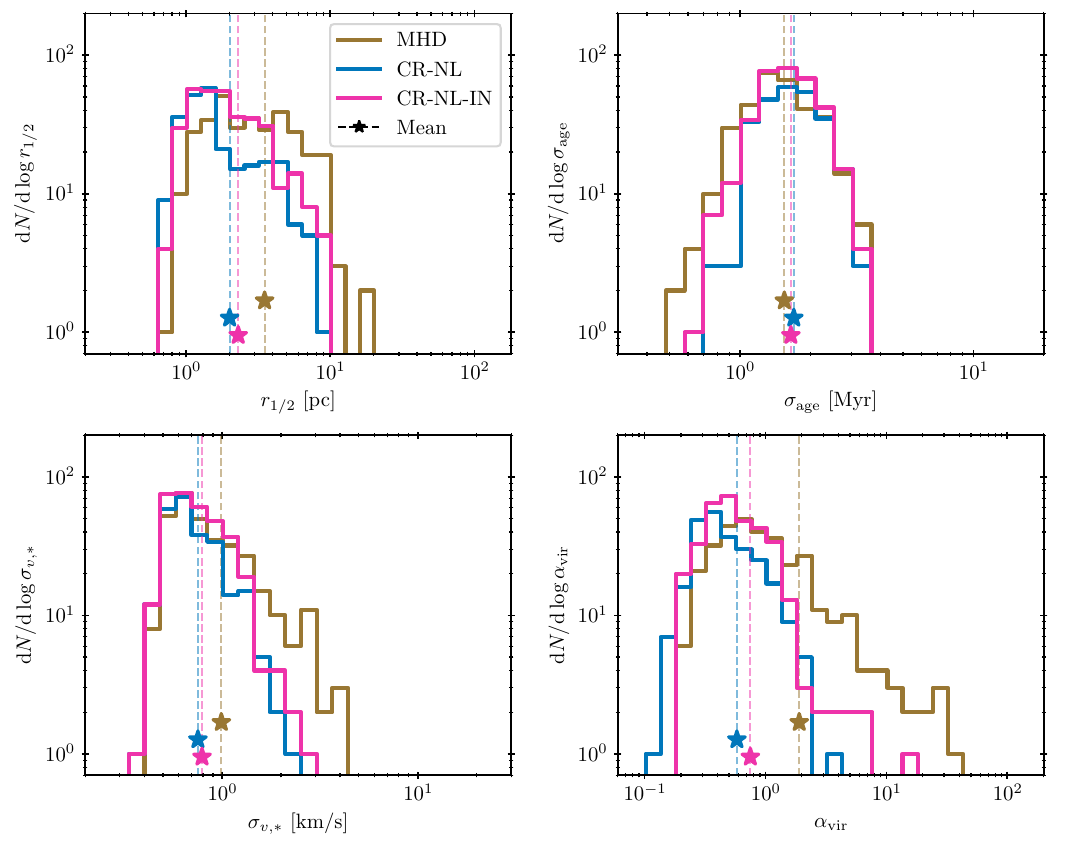}
\caption{Distributions of properties of young star clusters ($t_\mathrm{age}\simeq10\;\mathrm{Myr}$, $M>10^{3}\;\mathrm{M}_\odot$). We show the distribution of half mass radii $r_{1/2}$ (top left), stellar age standard deviation $\sigma_\mathrm{age}$ (top right), one-dimensional velocity dispersion $\sigma_{\velsym,*}$ (bottom left), and virial parameter $\alpha_\mathrm{vir}$ (bottom right). The mean value for each distribution is marked with a vertical dashed line and a $\star$, with an arbitrary vertical offset to prevent visual overlap. The star clusters in these simulations have physically reasonable properties. The \mhdX has clusters with larger $r_{1/2}$ and $\sigma_{\velsym,*}$ than the clusters in two \crmhd cases. Young star clusters in the \mhdX case are, on average, unbound.
\label{fig:fourplot_combo_props}}
\end{figure*}

In this section, we characterize the physical properties of the young star clusters. The properties of young star clusters should depend on the molecular cloud environments from which they form. Because our previous work showed that CRs change the amount of star-forming gas in the ISM \sttfp, we expect that CRs may affect the properties of star-forming molecular clouds. Any effects will be encoded in the properties of young star clusters.

For each young star cluster with mass greater than $10^{3}\;\mathrm{M}_\odot$, we calculate the half-mass radius $r_{1/2}$, the mass-weighted standard deviation of the stellar age distribution $\sigma_{\mathrm{age}}$, the \mbox{1-dimensional} velocity dispersion of the stellar velocities $\sigma_{\velsym,*}$ (calculated as the 3-dimensional velocity dispersion multiplied by $3^{-1/2}$), and finally the virial parameter $\alpha_\mathrm{vir}$, defined as
\begin{equation} 
   \label{eq:virial_parameter}
    \alpha_\mathrm{vir} = \frac{5 r_{1/2}\, \sigma_{\velsym,*}^2}{GM}.
\end{equation}
We bin these values by the log of the respective quantity (i.e., calculating $\mathrm{d}N/\mathrm{d}\log r_{1/2}$). The distributions and averages for $r_{1/2}$, $\sigma_{\mathrm{age}}$, $\sigma_{\velsym,*}$, and $\alpha_\mathrm{vir}$ are shown in Figure~\ref{fig:fourplot_combo_props}.

First, focusing on the top left panel of Figure~\ref{fig:fourplot_combo_props}, we see that young star clusters in each case generally have half-mass radii around $1-10\;\mathrm{pc}$. This is consistent with the observed distribution of radii for star clusters of masses $10^{3}\;\mathrm{M}_\odot-10^{4.5}\;\mathrm{M}_\odot$ in nearby galaxies \citep{brown_radii_2021} and the radii of nearby open clusters \citep{trumpler_preliminary_1930}. Although the mean radius of star clusters in the \mhdX case is larger than the means in the \crmhd cases, both values are physically plausible. Another important factor to note is that the distance at which the softened gravitational potential becomes Newtonian is $2.8\;\mathrm{pc}$, larger than the mean radii in both \crmhd cases. This is discussed in Appendix~\ref{sec:app_discuss_limitations}.

Next, we look at $\sigma_\mathrm{age}$ distributions in the top right panel of Figure~\ref{fig:fourplot_combo_props}. Star clusters in each of the three cases tend to have age distributions with a width of $1-3\;\mathrm{Myr}$, consistent with previous cloud-scale simulations \citep[e.g.,][]{grudic_when_2018} and observations of Milky Way young star clusters \citep{longmore_formation_2014}. We interpret this to be a result of SN feedback in our simulations. Star formation in these simulated clusters occurs in a brief $2-6\;\mathrm{Myr}$ epoch that is disrupted by the onset of early SN feedback. This is consistent with our minimum stellar age for SN ($3\;\mathrm{Myr}$). This idea of the formation epoch ended by SNe is explored further in Appendix~\ref{sec:app_formation_epoch}. Because there are no significant differences between the cases, we infer that CRs do not disrupt star formation in still-forming embedded clusters, as suggested in Section~\ref{sec:stellar_feedback_methods}. Rather, in our idealized simulations, star formation is truncated by the onset of SNe. In a more realistic simulation with early stellar winds and direct radiative feedback from massive stars, it is feasible that SN would only play a secondary role in disrupting star formation in a cluster \citep[e.g.,][]{rathjen_silcc_2021, smith_efficient_2021}.

The $\sigma_{\velsym,*}$ distributions in the bottom left panel of Figure~\ref{fig:fourplot_combo_props} show differences between the physics cases. The average velocity dispersions in the \crmhd cases are approximately $0.8\;\mathrm{km}\;\mathrm{s}^{-1}$, whereas the average velocity dispersion in the \mhdX case is $1.0\;\mathrm{km}\;\mathrm{s}^{-1}$. The \mhdX case also has a more significant tail of high-velocity-dispersion star clusters. Similar to the radii, all cases are broadly consistent with observations, and no physical constraints can be derived from these distributions of velocity dispersions.

Finally, looking at the distributions of $\alpha_\mathrm{vir}$ in the bottom right panel of Figure~\ref{fig:fourplot_combo_props}, we see the most significant disparities between the three cases. The \mhdX case produces star clusters with higher average virial parameters than star clusters in the \crmhd cases, with a significant tail of clusters with high ($>2$) virial parameters. The distributions of $\alpha_\mathrm{vir}$ are naturally explained by the distributions of $r_{1/2}$ and $\sigma_{\velsym,*}$. Because $r_{1/2}$ and $\sigma_{\velsym,*}$ are largest on average in the \mhdX case, and because these two terms are in the numerator of $\alpha_\mathrm{vir}$, the values of $\alpha_\mathrm{vir}$ are largest for clusters in the \mhdX case.

The distribution of virial parameters could potentially be sensitive to numerical and resolution effects, changes in the star formation prescription, and the inclusion of additional feedback mechanisms. Numerical and resolution effects are likely subdominant, as argued in Appendix~\ref{sec:app_discuss_limitations}. As discussed in Section~\ref{sec:cfe_subsub}, a simultaneous decrease in the star formation density threshold and $\epsilon_\mathrm{ff}$ would allow stars to form at lower densities, where gas is less likely to be gravitationally bound and virialized \citep[c.f.,][]{hislop_challenge_2022}. Direct radiative feedback may also decrease the fraction of clusters that are bound at formation \citep{hislop_challenge_2022}. We note that this presented work is intended to be a parameter study of CR feedback, and that the importance of these results is the differences between the cases, and therefore the contribution of CRs to the results.

Notably, the two \crmhd cases have average virial parameters below $1$, implying that these star clusters are bound on average. In contrast, the \mhdX case has star clusters with virial parameters above $1$ on average, meaning that these star clusters are unbound on average. In the next section, we investigate differences in molecular clouds that could result in differences in star cluster virial parameters.


\section{Impact of CRs}
\label{sec:impact_of_crs}

\begin{figure}
\includegraphics{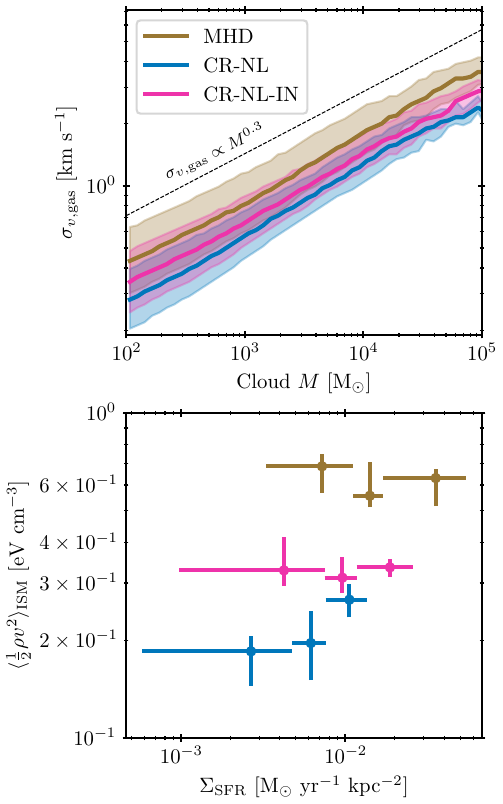}
\caption{\textit{Top}: Molecular cloud gas velocity dispersions $\sigma_{\velsym,\mathrm{gas}}$ as a function of the cloud mass $M$, where cloud identification is described in Section~\ref{sec:impact_of_crs}. We include the median $\sigma_{\velsym,\mathrm{gas}}$ at a given $M$, as well as the 25$^\mathrm{th}$ to 75$^{\mathrm{th}}$ percentiles in a given mass bin. Clouds in each case follow the apparent relation $\sigma_{\velsym,\mathrm{gas}}\propto M^{0.3}$ (included as a dashed line), implying that the transfer of kinetic energy to small scales is unchanged by CRs. However, the normalization of the relation is different, signaling a difference in total kinetic energy between the three cases. \textit{Bottom}: Volume-weighted kinetic energy density in the ISM, split by $\Sigma_\mathrm{SFR}$. We repeat the same SFRD binning process from Section~\ref{sec:environmental_dependence}, and display 25$^\mathrm{th}$ to 75$^{\mathrm{th}}$ percentiles of time-variability as error bars. The turbulent energy budget of the ISM is set by the overall environment, rather than the instantaneous SFRD.
\label{fig:cloud_velocity_dispersion_evolution}}
\end{figure}

Young star clusters inherit their ``boundedness'' from their progenitor molecular clouds \citep[e.g.,][]{kruijssen_fraction_2012} through their radii and velocity dispersions at the time of formation. Therefore, we seek to characterize the impact of CRs on star cluster formation by looking at differences in star-forming molecular clouds between the three cases.

We use the cloud-finding algorithm from \citet{weber_crexit_2025}. This algorithm identifies groups of gas cells using their connectivity in the Voronoi mesh. In our case, we run the cloud-finder for 17 different density thresholds (equally logspaced from $n_\mathrm{H}\geq10^{1.6}\;\mathrm{cm}^{-3}$ to $n_\mathrm{H}\geq10^{3.2}\;\mathrm{cm}^{-3}$), and consolidate the identified clouds into a single sample. We run this on all snapshots from $t=112-250\;\mathrm{Myr}$, capturing clouds at different evolutionary stages and identifying substructured cloud interiors. Analyzing every snapshot necessarily means that we view the same clouds multiple times. This should not introduce any meaningful systematic biases if we simply consider this to be a time-averaged sample (e.g., representative of an average single observation). After building this sample, we bin these clouds by their mass. Velocity dispersion should monotonically increase with cloud mass due to the scale-dependence of the turbulent cascade in the ISM \citep{larson_turbulence_1981}.

We show the cloud gas velocity dispersion as a function of cloud mass in Figure~\ref{fig:cloud_velocity_dispersion_evolution} (top panel). We include the median cloud velocity dispersion, as well as 25$^\mathrm{th}$ to 75$^{\mathrm{th}}$ percentiles by number. The three cases form power laws with approximate scaling $\sigma_{\velsym,\mathrm{gas}} \propto M^{0.3}$, but with slightly different normalizations. If CRs were to dynamically affect the collapse of molecular clouds, we would predict a mass/scale dependence, and consequently a different slope for the $\sigma_{\velsym,\mathrm{gas}} - M$ relations in the top panel Figure~\ref{fig:cloud_velocity_dispersion_evolution}. Instead, we see the same slopes but different normalizations, suggesting that CRs may not have a dynamical impact on molecular clouds.

This is not a fully unexpected result. \citet{fitz_axen_suppressed_2024} performed high-resolution simulations of the collapse of single molecular clouds, and found that CRs had little impact on the velocity dispersion of the cloud during collapse. Additionally, \citet{kjellgren_dynamical_2025} showed that CRs establish gradients on scales larger than the sizes of molecular clouds, implying that CRs cannot inhibit the collapse of dense molecular clouds. We suggest that the most important factor is the natal velocity dispersion of the cloud, here reflected by the normalization of the line in Figure~\ref{fig:cloud_velocity_dispersion_evolution}, top panel.

We can explain this through simplified arguments. If we assume that a) clouds are supported by velocity dispersion, b) clouds in a single case all share the same virial parameter, and c) clouds exist at approximately the same density, then the virial parameter in Equation~\ref{eq:virial_parameter} can be re-arranged to show that $\sigma_\mathrm{v}\propto M^{1/3}$. In Figure~\ref{fig:cloud_velocity_dispersion_evolution}, we see that each case roughly follows the relation $\sigma_{\velsym,\mathrm{gas}}\propto M^{0.3}$ (displayed as a dashed line in Figure~\ref{fig:cloud_velocity_dispersion_evolution}). Additionally, the relation $\sigma_{\velsym,\mathrm{gas}}\propto M^{0.3}$ has been previously identified in the analysis of simulated clouds without CR feedback \citep{banerjee_clump_2009}. Therefore, we can infer that the transfer of kinetic energy to smaller scales \citep[e.g.,][]{larson_turbulence_1981}\footnote{We do not include the thermal component of the velocity dispersion in our calculation, preventing a direct comparison with \citet{larson_turbulence_1981} or some other simulated work \citep[e.g.,][]{ni_life_2025}.} is not affected by CRs. Rather, the total kinetic energy budget available to clouds must be different between the three cases, explaining the differences in the normalizations in Figure~\ref{fig:cloud_velocity_dispersion_evolution}, top panel.

As reinforcement for this idea, we investigate the ISM kinetic energy density as a function of the SFRD. If cloud velocity dispersion is indeed set by the state of the ISM, the three cases should have notably distinct values for ISM kinetic energy density, independent of the instantaneous SFRD. We investigate this by splitting snapshots using the same SFRD binning process described in Section~\ref{sec:cluster_identification}. We calculate the volume-weighted average of the kinetic energy density $\langle\frac{1}{2} \rho \velsym^2\rangle$ for all gas cells in the ISM ($|z|<200\;\mathrm{pc}$), where $\velsym$ is evaluated in the lab frame. We report the time-averaged value and $25^{\mathrm{th}}$ and $75^{\mathrm{th}}$ percentiles of fluctuations of the kinetic energy density in Figure~\ref{fig:cloud_velocity_dispersion_evolution} (bottom panel).

The three runs have ISM kinetic energy densities that are almost entirely distinct from each other, regardless of the SFRD window in which this is calculated. The \mhdX case has gas with the highest ISM kinetic energy densities, followed by the \inX case and the \nlX case. This trend mirrors the trend of the average star formation rates for these cases. Only the \nlX demonstrates any positive correlation between instantaneous SFRD and ISM kinetic energy densities; however, the ISM kinetic energy densities in this case are still systematically lower than those in the other cases, suggesting that this correlation does not disprove our interpretation. We conclude that the kinetic energy budget of the ISM is a product of long-term trends in SFRD, rather than instantaneous SFRD. SNe occur over a wide variety of time-delays, producing a SN rate that varies less with time than the SFR. Instead, the SN rate (and consequent turbulent energy injection rate) depends on the long-term SFRD. When gas from the volume-filling phase of the ISM collapses into clouds, the gas retains its kinetic energy via compression and turbulent cascade, and therefore retains information of the state of the ISM and turbulent energy supply. As such, the velocity dispersions of clouds and star clusters in our simulations are most significantly determined by the long-term behavior of the SFRD.

Note that this behavior (higher SFRD means less bound clusters) should not be extrapolated different galactic environments with different surface densities. Each case here shares the same initial gas mass surface density; therefore, differences in SFRD correspond to differences in kinetic energy supplied per unit gas mass, resulting in the differences in ISM kinetic energy and therefore velocity dispersions. Although these three cases show a negative correlation between $\Gamma$ and $\Sigma_\mathrm{SFR}$ for a single gas mass surface density with differing physics, the overall trend between $\Gamma$ and $\Sigma_\mathrm{SFR}$ across different surface densities is likely still positive (as in observed datasets). This is an example of Simpson's paradox \citep{simpson_interpretation_1951}---subsets of data can exhibit reversed trends when compared to the entire dataset.

Incidentally, the difference in cloud velocity dispersions between cases explains the differences in star cluster velocity dispersions and star cluster radii. When the molecular clouds form star clusters in the \mhdX, they begin with a larger velocity dispersion than those in the \crmhd cases. If a cluster is formed unbound (e.g., $\alpha_\mathrm{vir}>1$), it will expand over time. Consequently, unbound star clusters evolve for $\sim10\;\mathrm{Myr}$ and expand beyond their natal radius (the radius of the star-forming core, set by the mass and star-formation density threshold, assuming a spherical shape for star-forming cores). Therefore, initially unbound clusters will have both larger radii and larger velocity dispersions after $10\;\mathrm{Myr}$. As seen in Figure~\ref{fig:fourplot_combo_props}, the star clusters in the \mhdX case have larger velocity dispersions and radii than the \crmhd cases, translating to larger virial parameters in the \mhdX case.

Overall, we find that CRs under the theory of self-confinement do not seem to affect the evolution of individual dense clouds after they have already formed. Rather, the impact of CRs is to change the overall distribution of gas mass in the ISM \sttfp, reducing the SFR and consequently the equilibrium turbulent energy density due to SNe. Future work will involve a more thorough investigation into precisely how CRs dynamically affect the ISM and reduce SFRs (Sike et. al. in prep).



\section{Summary and Conclusions}
\label{sec:conclusions}

In this paper, we investigate the star clusters formed in the simulations from \sttf. These simulations adopt the tallbox approach with solar neighborhood-like conditions, varied implementations of CRs under the theory of self-confinement, and the \textsc{Crisp} feedback framework. Most importantly, these are the first simulations of star cluster formation to simultaneously include dynamically-coupled CRs, a multiphase ISM environment, and individual star particles representative of massive stars. We compare our simulated star clusters with observed star cluster populations to test our model, providing valuable information for future simulations of galactic feedback. We summarize our findings below.

\begin{enumerate}

    \item We confirm the existence of star clusters in these simulations, despite our idealized setup. We find these star clusters to have physically plausible mass functions with power law slopes and CFEs that are positively correlated with the SFRD, as found in previous literature. Adding CRs shifts our results within observationally-permitted regions, along observed trends.
    
    \item In all three of our simulation cases, approximately half of SNe occur in diffuse ambient gas. This result does not depend on the inclusion of CRs, despite the fact that CRs significantly change the volume fraction of the diffuse ISM.

    \item Star clusters in the \mhdX case have higher half mass radii and velocity dispersions than star clusters in the \crmhd cases. Consequently, star clusters in the \mhdX case are unbound ($\alpha_\mathrm{vir}>1$) on average. The \mhdX case has the highest star formation rates, translating to a larger turbulent energy budget in the ISM and higher velocity dispersions in star-forming molecular clouds.
\end{enumerate}

In summary, the primary effect of dynamically-coupled CRs is to reduce the SFR. The overall properties of our star cluster populations are consistent with observations, regardless of the inclusion of CRs.


\begin{acknowledgements}
We thank the anonymous referee for their comments that helped improve the clarity and justification of this manuscript. B.S. and M.R. thank the Leibniz Institute for Astrophysics Potsdam (AIP) for its hospitality during their stay, where part of this work was performed. B.S. thanks Sean D. Johnson, Colin Holm-Hansen, and Aster G. Taylor for constructive discussions. 

M.R. acknowledges support from the National Aeronautics and Space Administration grant ATP 80NSSC23K0014 and the National Science Foundation Collaborative Research Grant NSF AST-2009227. O.Y.G. and Y.C. were supported in part by National Aeronautics and Space Administration through contract NAS5-26555 for Space Telescope Science Institute programs HST-AR-16614 and JWST-GO-03433. M.W., T.T., and C.P. acknowledge support from the European Research Council under ERC-AdG grant PICOGAL-101019746. This work was supported by the North-German Supercomputing Alliance (HLRN) under project bbp00070. This material is based upon work supported by the National Science Foundation Graduate Research Fellowship Program under Grant No. DGE 2241144. Any opinions, findings, and conclusions or recommendations expressed in this material are those of the author(s) and do not necessarily reflect the views of the National Science Foundation. 

This work used the Delta system at the National Center for Supercomputing Applications through allocation PHY240111 from the Advanced Cyberinfrastructure Coordination Ecosystem: Services \& Support (ACCESS) program, which is supported by National Science Foundation grants Nos. 2138259, 2138286, 2138307, 2137603, and 2138296 \citep{boerner_access_2023}. This research was supported in part through computational resources and services provided by Advanced Research Computing (ARC)--a division of Information and Technology Services (ITS) at the University of Michigan, Ann Arbor.
\end{acknowledgements}



\begin{appendix}

\section{Discussion of Limitations}
\label{sec:app_discuss_limitations}

In this section, we qualitatively discuss the limitations imposed by the idealized setup used for this work. Specifically, we focus on limitations not extensively covered in the body of this work. This discussion is not integral to the understanding of the primary results of this work.

Due to the idealized ``tallbox'' setup, we do not capture the effects of global galactic structure like shear or spiral arms. These can affect the statistical star formation properties \citep[e.g.,][]{kim_three-phase_2017}, lower the maximum mass for molecular clouds \citep[e.g.,][]{toomre_gravitational_1964, reina-campos_unified_2017}, and provide additional environments to study star clusters within a single galaxy \citep[e.g.,][]{messa_young_2018, grudic_great_2023}. Additionally, the tallbox setup limits the maximum gas mass available for molecular cloud formation. This could artificially limit the maximum cluster mass in our simulations to a mass lower than what would be expected for the solar neighborhood. Other geometric limitations of the tallbox setup are discussed in \sttf.

Because we do not resolve the IMF or include any subgrid prescription for star particles of different masses, we miss out on several physical processes \citep{smith_sensitivity_2021}. We do not capture the time-dependence of the formation of differently-massed stars \citep{grudic_does_2023}. If massive stars are able to form first, this could lead to more rapid feedback and an earlier truncation of star formation in a star cluster, resulting in lower star cluster masses overall. Competitive accretion \citep{bonnell_competitive_2001} could lead to a greater concentration of high-mass stars towards the center of a star cluster, altering the clustering of SNe and the location/ambient media of the first SNe of a cluster. The variable energy-injection of SNe based on the progenitor star masses can also have an impact on the results of SN feedback \citep{gutcke_lyra_2021, smith_sensitivity_2021}. We also do not include stellar winds or radiation feedback, which could alter the ambient media as well \citep{rathjen_silcc_2021} and truncate star cluster formation earlier than with SNe alone \citep{gatto_silcc_2017}.

A potentially significant numerical limitation of our work is the gravitational softening length. We adopt a gravitational softening length of $1\;\mathrm{pc}$, implying that the gravitational potential for a star particle does not become Newtonian until $2.8\;\mathrm{pc}$ \citep{springel_e_2010}. Many star clusters have radii smaller than this number (see Figure~\ref{fig:fourplot_combo_props}, top left panel), suggesting that the bulk of star-star gravitational interactions within a star cluster are unresolved and heavily softened.

Rather surprisingly, the gravitational softening should not have a significant impact on the presented velocity dispersions of young star clusters. There are several reasons for this. Firstly, the total gravitational potential energy of the interior of a star cluster approximately depends on the mean separation between member stars. A star cluster with half mass radius $r_{1/2}$ will have interior member stars with average separations of order $r_{1/2}$. If $r_{1/2}$ is comparable to, or slightly less than the gravitational softening length, then the ``dilution`` of the gravitational potential due softening between each pairwise interaction only minimally affects the total potential (and thus the velocity dispersion). The majority of pairwise interactions occur over long-distances ($\Delta r \gtrsim r_{1/2}$) and are unsoftened. A fair amount of moderately-spaced ($\Delta r \lesssim r_{1/2}$) pairwise interactions are slightly softened, and a select few interactions ($\Delta r \ll r_{1/2}$) are heavily softened. The overall net effect is a relatively small reduction in the central gravitational potential of the cluster. Secondly, the impact of gravitational softening outside $r_{1/2}$ is minimal. For sufficiently non-compact clusters (e.g., $r_{1/2}$ is not substantially smaller than $2.8\;\mathrm{pc}$), member stars outside $r_{1/2}$ effectively experience a Newtonian potential. These outer stars, by definition, constitute half of the mass of the star cluster, and therefore determine half of the velocity dispersion. The ``error'' in the velocity dispersion due to stars within $r_{1/2}$ is therefore mitigated by a factor of $\sim1/2$, due to the relatively error-free outskirts. We note, however, that these arguments do not hold for star clusters with $r_{1/2} \lesssim 1\;\mathrm{pc}$. Fortunately, clusters with $r_{1/2}<1\;\mathrm{pc}$ are generally less massive than our mass cut of $10^{3}\;\mathrm{M}_\odot$, and excluded from our analysis.

We also expect the virial parameters and boundedness calculations to be robust. Despite the softened gravitational potential, the \textsc{Arepo} N-body integrator aims to conserve energy under dynamics \citep{springel_e_2010}. The total energy of a star cluster system should not depend on the form of the gravitational potential, assuming energy is conserved. The final boundedness of a star cluster is primarily determined by the initial energy budget of the collapsing molecular cloud, and the amount of energy lost during collapse, regardless of the form of the potential. Therefore, we do not expect these energies to be significantly impacted by gravitational softening.

Although our setup is oversimplified for all of the reasons outlined above (and in the main text), we still naturally produce star clusters with realistic mass functions and properties. This work demonstrates a promising success for the \textsc{Crisp} feedback model, particularly on a facet of star formation that was never intended to be analyzed. Observations of star clusters are a great metric for testing the parameters of our feedback model, and allow us to better understand the impact of our parameter choices such as our star formation density threshold and star formation efficiency, as well our CR transport model. With the insights gained from this work, we will be able to improve future iterations of simulations, working towards a predictive model for galaxy formation and evolution \citep[see][for motivating discussion]{naab_theoretical_2017}.

\section{Discussion of FoF Parameters}
\label{sec:app_discuss_fof}

\begin{figure*}
\includegraphics{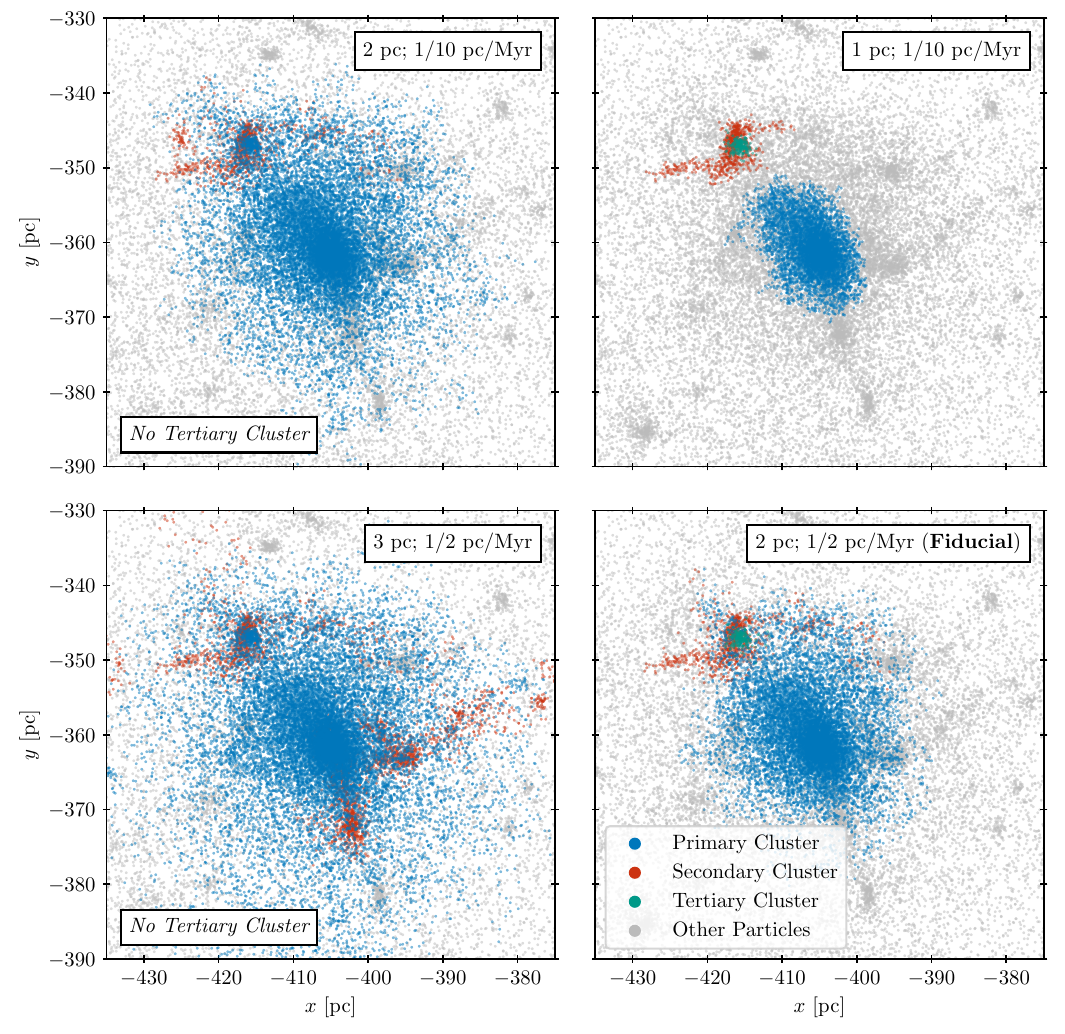}
\caption{Demonstration of the results of the FoF star cluster finder for the \mhdX case at $t=250.31\;\mathrm{Myr}$, focusing on three neighboring star clusters and applying four different sets of parameters. We include the space and time separation $\Delta w$ and the time-to-space conversion $u_0$ in the top right of each panel. FoF finders with the non-fiducial parameter sets have difficulty identifying the three clusters. With the fiducial set of parameters, we identify a large number of members for the primary cluster, identify the secondary cluster as a localized object, and identify the tertiary cluster as a separate object.
\label{fig:fof_comparison_params_appendix}}
\end{figure*}

Our FoF algorithm has two free parameters: the time and space separation $\Delta w$, and the time-to-length conversion factor $u_0$. We adopt $\Delta w = 2\;\mathrm{pc}$ and $u_0=1/2\;\mathrm{pc}/\mathrm{Myr}$ for our fiducial parameters. These choices are a compromise between lenient member identification for isolated clusters, and strict delineation between neighboring clusters. To demonstrate the effect of these parameters, we test several parameter sets on a difficult system of neighboring clusters found in the \mhdX case. This system contains a large primary cluster, a secondary tidally-stripped cluster, and a small overlapping tertiary cluster, all contained within a $\sim(50\;\mathrm{pc})^{3}$ region. We show the results of the FoF cluster-identification algorithm with different parameter sets on this system in Figure~\ref{fig:fof_comparison_params_appendix}.

With the first parameter choice $\Delta w = 2\;\mathrm{pc}; \;u_0=1/10\;\mathrm{pc}/\mathrm{Myr}$ (Figure~\ref{fig:fof_comparison_params_appendix}, top left panel), we see that the primary cluster and secondary cluster are successfully identified; however, the tertiary cluster is lumped in with the primary cluster. This suggests that the linking length is too large, or the time-space conversion is too generous. In the top right panel of Figure~\ref{fig:fof_comparison_params_appendix}, we test a stricter linking length $\Delta w = 1\;\mathrm{pc}$, but maintain $u_0=1/10\;\mathrm{pc}/\mathrm{Myr}$. The three clusters are successfully identified as unique objects with this parameter choice. Unfortunately, due to the short linking length, the primary cluster is small, and the outskirts are artificially truncated. We test a larger linking length $\Delta w = 3\;\mathrm{pc}$ with a stricter time-space conversion $u_0=1/2\;\mathrm{pc}/\mathrm{Myr}$ and show the results in the bottom right panel of Figure~\ref{fig:fof_comparison_params_appendix}. In this case, similar to the first parameter set, the tertiary cluster is incorrectly identified as part of the primary cluster. Additionally, the secondary cluster is spuriously over-extended, demonstrating that the linking length is overly generous. Finally, in Figure~\ref{fig:fof_comparison_params_appendix} (bottom right panel), we show the results for the fiducial parameter choices $\Delta w = 2\;\mathrm{pc}; \;u_0=1/2\;\mathrm{pc}/\mathrm{Myr}$. In this case, the primary, secondary, and tertiary cluster are successfully identified as separate objects, and neither cluster shows obvious signs of contamination in the outskirts. Our fiducial parameter choice is the most successful parameter choice in handling this complicated system of neighboring clusters.

The fiducial choice of parameters does not represent a ``correct'' cluster-identification algorithm, nor does it represent a unique solution. The average interparticle spacing of a cluster of a given mass will depend on the average mass of a star particle. In our case, this is $\sim10\;\mathrm{M}_\odot$, which is larger than expected for real star clusters. Additionally, the age selection criterion will likely depend on the cluster formation rate and the likelihood of two neighboring independent clusters forming at the same time. We emphasize that the adopted parameters satisfy the needs for this work in identifying individual self-gravitating clusters, with only a moderately strict membership criterion. The central conclusions of this work are not affected by small changes to these parameters. 

We additionally tested higher values of $u_0$, and we found that these FoF methods failed to identify bound members in the outskirts of clusters, owing to the additional strictness in the time component. A larger $u_0$ creates a more effective ``spread'' in the age distribution processed by the FoF catalog, causing the time component to quickly dominate the linking length in the outskirts of clusters.

\section{Formation Epoch}
\label{sec:app_formation_epoch}

The distributions of $\sigma_\mathrm{age}$ values in Figure~\ref{fig:fourplot_combo_props} (upper right) are very consistent between the three physics cases. This consistency suggests a common physical mechanism that limits the formation epoch of a star cluster, such as SN feedback. Here, we investigate the SFR histories and SN rate histories of clusters to determine if the distributions of $\sigma_\mathrm{age}$ values are explained by SN feedback.

Tracking the star formation history of an individual young star cluster is nontrivial. Star clusters can form as adjacent clumps of stars (see star-forming gas cells in Figure~\ref{fig:pretty_star_plot}), and may eventually overlap or spread apart. The star formation history of a single, unique, star cluster is ill-defined. We avoid this problem by analyzing each star formation event individually, using a ``stars-first'' approach, governed by the following inequality,
\begin{equation}
    {\Delta \xi } \equiv \left|\bm{r}_\mathrm{c,com} + \bm{\velsym}_\mathrm{c,com} \left(t_\star-t_\mathrm{snap} \right) -\bm{r}_\star\right| < 2^{1/3}\,r_{\mathrm{c},1/2}.
\end{equation}
For each star formation event occurring at time $t_\star$ and location $\bm{r}_\star$, we find the nearest snapshot in time $t_\mathrm{snap}$ and extrapolate the motions of all clusters to first-order using their center-of-mass $\bm{r}_\mathrm{c,com}$ and center-of-mass velocity $\bm{\velsym}_\mathrm{c,com}$. The distance between the extrapolated center-of-mass and the star formation site must be less than $2^{1/3}$ times the half mass radius of the star cluster $r_{\mathrm{c},1/2}$. This distance $2^{1/3}\,r_{\mathrm{c},1/2}$ would be the full radius of a cluster if it were a uniform sphere of constant density, where half of the mass was contained inside $r_{1/2}$ and half outside. For every cluster satisfying this criterion, the stellar mass of the newly-formed star particle is binned at $t_\star-t_{60}$, where $t_{60}$ is the time at which the cluster contained only $60\;\mathrm{M}_\odot$ (see Section~\ref{sec:cluster_identification} for further discussion of $t_{60}$). Therefore, $t_\star-t_{60}$ is an approximation for the age of the star cluster at the time of star formation, and the amount of mass in a given timebin describes the SFR as a function of star cluster age.

This ``stars-first'' binning process is effectively equivalent to stacking the star formation histories of each cluster. Each star particle's mass is binned into the total star formation history. We therefore produce an aggregate cluster star formation history for each case. This binning method has several benefits. Star particles are binned at their formation time with their formation mass. This prevents underestimation of the SFH due to mass-loss events such as SNe or cluster disruption. Additionally, we avoid ambiguity in the definition of a single star cluster through time. This can be especially important during the early epoch of star cluster formation, when a handful of star-forming molecular cloud cores may contribute stars to the same star cluster, or when an unbound star cluster rapidly dissolves into several open clusters. For the above reasons, we adopt this stars-first binning approach.

\begin{figure*}
\includegraphics{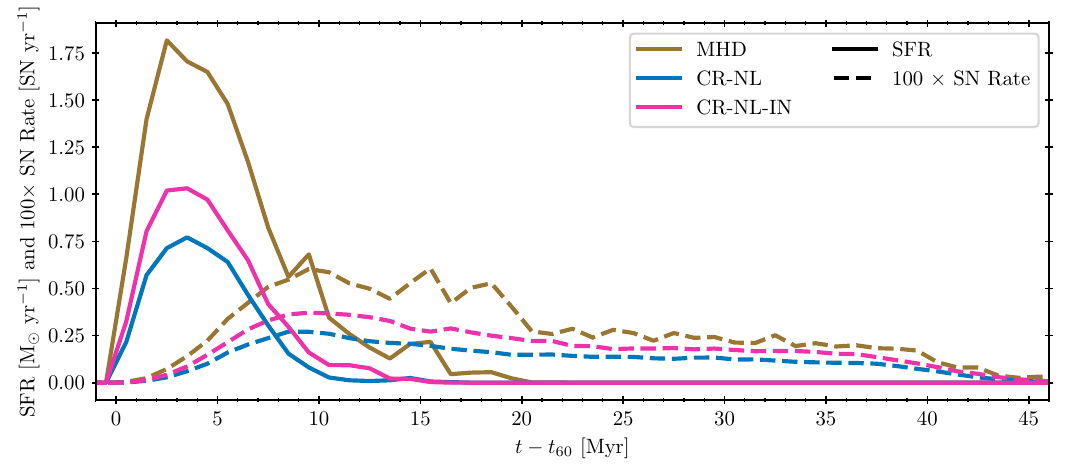}
\caption{Stacked SFR histories and SN rate histories for all star clusters in each of the three runs, relative to $t=t_{60}$. The membership criterion is defined in Appendix~\ref{sec:app_formation_epoch}, and $t_{60}$ is described in Section~\ref{sec:cluster_identification}. The SN rate is multiplied by a factor of $100$ to produce similar normalization between the SFR and SN rate histories. The bulk of cluster star formation occurs between $0$ and $\sim10\;\mathrm{Myr}$, followed by an extended epoch of SNe from $\sim10\;\mathrm{Myr}$ to $\sim40\;\mathrm{Myr}$. The onset of SNe appears to coincide with a decrease in the SFR, suggesting a link between SN feedback and the truncation of star formation in these simulated clusters. Note that the physical relevance of this result is limited, as radiative feedback and stellar winds would likely contribute to an earlier truncation of star formation. Rather, CR feedback does not seem to impact the duration of the epoch of star formation.
\label{fig:sfr_snrate_formation_epoch}}
\end{figure*}

We show the stacked SFR histories and SN rate histories (calculated with the binning process described above) in Figure~\ref{fig:sfr_snrate_formation_epoch}. We display the SN rate with an extra factor of $100$ to produce similar normalizations (and for readability) between the SFR and SN rate histories, because our model produces $1$ SN per $100\;\mathrm{M}_\odot$ of stellar mass formed (on average).

The SFR and SN rate histories in the three cases follow the same general shape. For the first $2-3\;\mathrm{Myr}$ after $t_{60}$, the SFR increases rapidly. Then, the SN rate begins to increase. This is because the minimum star particle age allowed for a SN is $3\;\mathrm{Myr}$. The SFRs continue to increase, but begin to flatten out. The SFR history in each case peaks at around $3-4\;\mathrm{Myr}$ after $t_{60}$, coinciding with the continued increase of the SN rates. This peak is followed by a rapid decline in the SFR until $t-t_{60}\approx8-10\;\mathrm{Myr}$, after which the SFR slowly tapers to zero.

The relationship between the stifling of the SFRs and the onset of SNe could be an indication of feedback behavior. In the absence of other feedback mechanisms such as stellar winds and radiation, the star-forming epoch of a star cluster can be truncated by SNe \citep[e.g.,][]{ni_life_2025}. Because we do not include these other feedback channels, SNe are the only immediate feedback mechanism. Therefore, we determine that SN feedback is likely the cause of the truncation of star formation in clusters in our simulations, rather than the depletion of available gas via a complete conversion into stars. The duration of the star-forming epoch does not seem to depend on the inclusion of CR feedback. Thermal and kinetic feedback from SNe alone are enough to halt star formation in clusters. These conclusions are limited to understanding the processes occurring in the simulations. In reality, radiation and stellar winds would likely contribute additional feedback to halt star formation.

After the SN rates' initial increase at $3\;\mathrm{Myr}$, they increase until a peak at $\sim10\;\mathrm{Myr}$, followed by an extended, $\sim30\;\mathrm{Myr}$-long decline. This is trivially due to the SN prescription, where star particles with an age between $3$ and $38\;\mathrm{Myr}$ can become a SN. After $t-t_{60}\approx45\;\mathrm{Myr}$, the SN rate has effectively dropped to zero, as even the youngest star particles are too old to become SNe.

The vertical ordering of the three cases in Figure~\ref{fig:sfr_snrate_formation_epoch} is consistent with the differing overall SFRs from the three cases. The \mhdX case forms the most stellar mass, and is therefore expected to form the most stellar mass in clusters as well. Additionally, the CFE scales positively with the SFRD, suggesting that the \mhdX case has the highest cluster formation rates as well. Because each case has a star formation epoch of approximately the same duration, the \mhdX case will therefore always have the highest star formation rate in clusters. The \inX and \nlX cases have lower SFRDs, and therefore lower normalizations in Figure~\ref{fig:sfr_snrate_formation_epoch}.

The SFR peaks in Figure~\ref{fig:sfr_snrate_formation_epoch} are clearly wider than even the highest values of $\sigma_\mathrm{age}$ from Figure~\ref{fig:fourplot_combo_props}. This is simply a statistical effect produced by the stacking of many clusters' SFR histories for Figure~\ref{fig:sfr_snrate_formation_epoch}. Individual star clusters likely experience peaks in star formation at different values of $t-t_{60}$, with different durations. When aggregated, the SFR histories of these clusters will produce an overall distribution with a width that is necessarily greater than the individual widths. It is therefore expected that the SFR history of any individual cluster likely consists of a thinner peak than the stacked peak in Figure~\ref{fig:sfr_snrate_formation_epoch}.

Another apparent feature is several bumps in the SFR and SN rate histories of the \mhdX case. Particularly, there are large bumps at $2-3$, $9-10$, $13-17$, and $16-20\;\mathrm{Myr}$ in the SFR histories. The last two bumps in the SFR histories are associated with bumps in the SN rate histories as well. We attribute this to artifacts in the ``stars-first'' approach for the SFR and SN rate history calculations. We determined that a pre-existing star cluster of age $\gtrsim15\;\mathrm{Myr}$ overlapped with a star-forming molecular cloud, causing the new star formation to be identified within the SFR history of the pre-existing star cluster. This additionally means that the new SNe from the just-formed star cluster are added into the SN rate history of the pre-existing star cluster, producing the bumps seen at $13-17$, and $16-20\;\mathrm{Myr}$. We do not attribute any physical significance to these bumps. Additionally, the FoF algorithm successfully identifies the pre-existing star cluster and the newly formed star cluster as separate objects (See Appendix~\ref{sec:app_discuss_fof}).

\section{Data Table}
\label{sec:app_data_table}

We report the data used for Figures~\ref{fig:mass_function},~\ref{fig:alpha_vs_sfrd}, and~\ref{fig:gamma_vs_sfrd} in Table~\ref{tab:data_table_appendix}. Additionally, we include $\Gamma$ values for the full cluster sample in each case, including both lower bound cluster mass cuts ($1000\;\mathrm{M}_\odot$ and $5000\;\mathrm{M}_\odot$).

\begin{table*}
    \centering
    \begin{tabular}{l c c c c c}
    \hline
    \hline
    \noalign{\smallskip}
    Case & $\Sigma_\mathrm{SFR}$ range [$10^{-2}\,\mathrm{M}_\odot$ yr$^{-1}$ kpc$^{-2}$] & $N> 10^{3}\,\mathrm{M}_\odot$ & $\alpha_\mathrm{PL}$ & $\Gamma_{1000}$ [\%] & $\Gamma_{5000}$ [\%] \\
    \noalign{\smallskip}
    \hline
    \noalign{\smallskip}
    \mhdX & $0.330 - 5.396$ & 328 & $-2.08 \pm 0.06$ & $30.9$ & $14.0$ \\
    \noalign{\smallskip}
    -- low & $0.330 - 1.115$ & 64 & $-2.53 \pm 0.17$ & $29.2$ & $6.48$ \\
    \noalign{\smallskip}
    -- medium & $1.115 - 1.708$ & 148 & $-2.20 \pm 0.08$ & $40.2$ & $10.8$ \\
    \noalign{\smallskip}
    -- high & $1.708 - 5.396$ & 116 & $-1.84 \pm 0.08$ & $26.6$ & $17.6$ \\
    \noalign{\smallskip}
    \hline
    \noalign{\smallskip}
    \nlX & $0.058 - 1.361$ & 253 & $-2.38 \pm 0.07$ & $57.5$ & $11.1$ \\
    \noalign{\smallskip}
    -- low & $0.058 - 0.476$ & 40 & $-2.49 \pm 0.17$ & $54.5$ & $4.39$ \\
    \noalign{\smallskip}
    -- medium & $0.476 - 0.759$ & 85 & $-2.50 \pm 0.13$ & $60.9$ & $11.2$ \\
    \noalign{\smallskip}
    -- high & $0.759 - 1.361$ & 128 & $-2.29 \pm 0.09$ & $56.5$ & $13.1$ \\
    \noalign{\smallskip}
    \hline
    \noalign{\smallskip}
    \inX & $0.097 - 2.567$ & 341 & $-2.28 \pm 0.05$ & $51.9$ & $13.1$ \\
    \noalign{\smallskip}
    -- low & $0.097 - 0.750$ & 59 & $-2.41 \pm 0.12$ & $49.8$ & $2.42$ \\
    \noalign{\smallskip}
    -- medium & $0.750 - 1.172$ & 122 & $-2.35 \pm 0.10$ & $54.0$ & $11.1$ \\
    \noalign{\smallskip}
    -- high & $1.172 - 2.567$ & 160 & $-2.21 \pm 0.08$ & $51.2$ & $17.6$ \\
    \noalign{\smallskip}
    \\
    \end{tabular}
    \caption{Cluster mass function results for the three cases, split between total, low SFRD, medium SFRD, and high SFRD bins. The $\Sigma_\mathrm{SFR}$ range column includes the lower and upper bounds of SFRD for the cluster sample. $\alpha_\mathrm{PL}$ is fit to the cluster mass function above $10^{3}\;\mathrm{M}_\odot$ (fit to a sample size of $N>10^{3}\;\mathrm{M}_\odot$ clusters), with errors calculated via bootstrapping, as in Section~\ref{sec:mass_function_sub}. The value for $\Gamma$ is calculated as the sum of bound cluster masses above a specific mass ($1000\;\mathrm{M}_\odot$ for $\Gamma_{1000}$, $5000\;\mathrm{M}_\odot$ for $\Gamma_{5000}$) divided by the total mass of stars formed in the respective SFRD bin, as in Section~\ref{sec:cfe_subsub}.
    \label{tab:data_table_appendix}}
\end{table*}

\end{appendix}

\bibliography{main}

\begin{thebibliography}{}
\expandafter\ifx\csname natexlab\endcsname\relax\def\natexlab#1{#1}\fi
\providecommand{\url}[1]{\href{#1}{#1}}
\providecommand{\dodoi}[1]{doi:~\href{http://doi.org/#1}{\nolinkurl{#1}}}
\providecommand{\doeprint}[1]{\href{http://ascl.net/#1}{\nolinkurl{http://ascl.net/#1}}}
\providecommand{\doarXiv}[1]{\href{https://arxiv.org/abs/#1}{\nolinkurl{https://arxiv.org/abs/#1}}}

\bibitem[{{Abrahamsson} {et~al.}(2007){Abrahamsson}, {Krems}, \&
  {Dalgarno}}]{abrahamsson_fine-structure_2007}
{Abrahamsson}, E., {Krems}, R.~V., \& {Dalgarno}, A. 2007, \apj, 654, 1171,
  \dodoi{10.1086/509631}

\bibitem[{{Adamo} {et~al.}(2015){Adamo}, {Kruijssen}, {Bastian}, {Silva-Villa},
  \& {Ryon}}]{adamo_probing_2015}
{Adamo}, A., {Kruijssen}, J.~M.~D., {Bastian}, N., {Silva-Villa}, E., \&
  {Ryon}, J. 2015, \mnras, 452, 246, \dodoi{10.1093/mnras/stv1203}

\bibitem[{{Adamo} {et~al.}(2011){Adamo}, {{\"O}stlin}, \&
  {Zackrisson}}]{adamo_probing_2011}
{Adamo}, A., {{\"O}stlin}, G., \& {Zackrisson}, E. 2011, \mnras, 417, 1904,
  \dodoi{10.1111/j.1365-2966.2011.19377.x}

\bibitem[{{Adamo} {et~al.}(2020{\natexlab{a}}){Adamo}, {Hollyhead}, {Messa},
  {Ryon}, {Bajaj}, {Runnholm}, {Aalto}, {Calzetti}, {Gallagher}, {Hayes},
  {Kruijssen}, {K{\"o}nig}, {Larsen}, {Melinder}, {Sabbi}, {Smith}, \&
  {{\"O}stlin}}]{adamo_star_2020}
{Adamo}, A., {Hollyhead}, K., {Messa}, M., {et~al.} 2020{\natexlab{a}}, \mnras,
  499, 3267, \dodoi{10.1093/mnras/staa2380}

\bibitem[{{Adamo} {et~al.}(2020{\natexlab{b}}){Adamo}, {Zeidler}, {Kruijssen},
  {Chevance}, {Gieles}, {Calzetti}, {Charbonnel}, {Zinnecker}, \&
  {Krause}}]{adamo_star_2020-1}
{Adamo}, A., {Zeidler}, P., {Kruijssen}, J.~M.~D., {et~al.} 2020{\natexlab{b}},
  \ssr, 216, 69, \dodoi{10.1007/s11214-020-00690-x}

\bibitem[{{Almeida} {et~al.}(2025){Almeida}, {Moitinho}, \&
  {Moreira}}]{almeida_open_2025}
{Almeida}, D., {Moitinho}, A., \& {Moreira}, S. 2025, \aap, 693, A305,
  \dodoi{10.1051/0004-6361/202451853}

\bibitem[{{Andersson} {et~al.}(2024){Andersson}, {Mac Low}, {Agertz}, {Renaud},
  \& {Li}}]{andersson_pre-supernova_2024}
{Andersson}, E.~P., {Mac Low}, M.-M., {Agertz}, O., {Renaud}, F., \& {Li}, H.
  2024, \aap, 681, A28, \dodoi{10.1051/0004-6361/202347792}

\bibitem[{{Annibali} {et~al.}(2011){Annibali}, {Tosi}, {Aloisi}, \& {van der
  Marel}}]{annibali_cluster_2011}
{Annibali}, F., {Tosi}, M., {Aloisi}, A., \& {van der Marel}, R.~P. 2011, \aj,
  142, 129, \dodoi{10.1088/0004-6256/142/4/129}

\bibitem[{{Armillotta} {et~al.}(2021){Armillotta}, {Ostriker}, \&
  {Jiang}}]{armillotta_cosmic-ray_2021}
{Armillotta}, L., {Ostriker}, E.~C., \& {Jiang}, Y.-F. 2021, \apj, 922, 11,
  \dodoi{10.3847/1538-4357/ac1db2}

\bibitem[{{Armillotta} {et~al.}(2024){Armillotta}, {Ostriker}, {Kim}, \&
  {Jiang}}]{armillotta_cosmic-ray_2024}
{Armillotta}, L., {Ostriker}, E.~C., {Kim}, C.-G., \& {Jiang}, Y.-F. 2024,
  \apj, 964, 99, \dodoi{10.3847/1538-4357/ad1e5c}

\bibitem[{{Armillotta} {et~al.}(2025){Armillotta}, {Ostriker}, \&
  {Linzer}}]{armillotta_energy-dependent_2025}
{Armillotta}, L., {Ostriker}, E.~C., \& {Linzer}, N.~B. 2025, \apj, 989, 140,
  \dodoi{10.3847/1538-4357/adea68}

\bibitem[{{Bakes} \& {Tielens}(1994)}]{bakes_photoelectric_1994}
{Bakes}, E.~L.~O., \& {Tielens}, A.~G.~G.~M. 1994, \apj, 427, 822,
  \dodoi{10.1086/174188}

\bibitem[{{Ballesteros-Paredes} {et~al.}(2020){Ballesteros-Paredes},
  {Andr{\'e}}, {Hennebelle}, {Klessen}, {Kruijssen}, {Chevance}, {Nakamura},
  {Adamo}, \& {V{\'a}zquez-Semadeni}}]{ballesteros-paredes_diffuse_2020}
{Ballesteros-Paredes}, J., {Andr{\'e}}, P., {Hennebelle}, P., {et~al.} 2020,
  \ssr, 216, 76, \dodoi{10.1007/s11214-020-00698-3}

\bibitem[{{Bally}(2016)}]{bally_protostellar_2016}
{Bally}, J. 2016, \araa, 54, 491, \dodoi{10.1146/annurev-astro-081915-023341}

\bibitem[{{Bally} {et~al.}(1987){Bally}, {Langer}, {Stark}, \&
  {Wilson}}]{bally_filamentary_1987}
{Bally}, J., {Langer}, W.~D., {Stark}, A.~A., \& {Wilson}, R.~W. 1987, \apjl,
  312, L45, \dodoi{10.1086/184817}

\bibitem[{{Banerjee} {et~al.}(2009){Banerjee}, {V{\'a}zquez-Semadeni},
  {Hennebelle}, \& {Klessen}}]{banerjee_clump_2009}
{Banerjee}, R., {V{\'a}zquez-Semadeni}, E., {Hennebelle}, P., \& {Klessen},
  R.~S. 2009, \mnras, 398, 1082, \dodoi{10.1111/j.1365-2966.2009.15115.x}

\bibitem[{{Bastian}(2008)}]{bastian_star_2008}
{Bastian}, N. 2008, \mnras, 390, 759, \dodoi{10.1111/j.1365-2966.2008.13775.x}

\bibitem[{{Bergin} \& {Tafalla}(2007)}]{bergin_cold_2007}
{Bergin}, E.~A., \& {Tafalla}, M. 2007, \araa, 45, 339,
  \dodoi{10.1146/annurev.astro.45.071206.100404}

\bibitem[{{Bertoldi} \& {McKee}(1992)}]{bertoldi_pressure-confined_1992}
{Bertoldi}, F., \& {McKee}, C.~F. 1992, \apj, 395, 140, \dodoi{10.1086/171638}

\bibitem[{{Blondin} {et~al.}(1998){Blondin}, {Wright}, {Borkowski}, \&
  {Reynolds}}]{blondin_transition_1998}
{Blondin}, J.~M., {Wright}, E.~B., {Borkowski}, K.~J., \& {Reynolds}, S.~P.
  1998, \apj, 500, 342, \dodoi{10.1086/305708}

\bibitem[{Boerner {et~al.}(2023)Boerner, Deems, Furlani, Knuth, \&
  Towns}]{boerner_access_2023}
Boerner, T.~J., Deems, S., Furlani, T.~R., Knuth, S.~L., \& Towns, J. 2023, in
  Practice and {{Experience}} in {{Advanced Research Computing}} (Portland OR
  USA: ACM), 173--176, \dodoi{10.1145/3569951.3597559}

\bibitem[{{Bonnell} {et~al.}(2001){Bonnell}, {Bate}, {Clarke}, \&
  {Pringle}}]{bonnell_competitive_2001}
{Bonnell}, I.~A., {Bate}, M.~R., {Clarke}, C.~J., \& {Pringle}, J.~E. 2001,
  \mnras, 323, 785, \dodoi{10.1046/j.1365-8711.2001.04270.x}

\bibitem[{{Bonnell} {et~al.}(2003){Bonnell}, {Bate}, \&
  {Vine}}]{bonnell_hierarchical_2003}
{Bonnell}, I.~A., {Bate}, M.~R., \& {Vine}, S.~G. 2003, \mnras, 343, 413,
  \dodoi{10.1046/j.1365-8711.2003.06687.x}

\bibitem[{{Boulares} \& {Cox}(1990)}]{boulares_galactic_1990}
{Boulares}, A., \& {Cox}, D.~P. 1990, \apj, 365, 544, \dodoi{10.1086/169509}

\bibitem[{{Brown} \& {Gnedin}(2021)}]{brown_radii_2021}
{Brown}, G., \& {Gnedin}, O.~Y. 2021, \mnras, 508, 5935,
  \dodoi{10.1093/mnras/stab2907}

\bibitem[{{Brown} \& {Gnedin}(2022)}]{brown_testing_2022}
---. 2022, \mnras, 514, 280, \dodoi{10.1093/mnras/stac1164}

\bibitem[{{Buck} {et~al.}(2019){Buck}, {Dutton}, \&
  {Macci{\`o}}}]{buck_observational_2019}
{Buck}, T., {Dutton}, A.~A., \& {Macci{\`o}}, A.~V. 2019, \mnras, 486, 1481,
  \dodoi{10.1093/mnras/stz969}

\bibitem[{{Buck} {et~al.}(2020){Buck}, {Pfrommer}, {Pakmor}, {Grand}, \&
  {Springel}}]{buck_effects_2020}
{Buck}, T., {Pfrommer}, C., {Pakmor}, R., {Grand}, R. J.~J., \& {Springel}, V.
  2020, \mnras, 497, 1712, \dodoi{10.1093/mnras/staa1960}

\bibitem[{{Calura} {et~al.}(2022){Calura}, {Lupi}, {Rosdahl}, {Vanzella},
  {Meneghetti}, {Rosati}, {Vesperini}, {Lacchin}, {Pascale}, \&
  {Gilli}}]{calura_sub-parsec_2022}
{Calura}, F., {Lupi}, A., {Rosdahl}, J., {et~al.} 2022, \mnras, 516, 5914,
  \dodoi{10.1093/mnras/stac2387}

\bibitem[{{Calura} {et~al.}(2025){Calura}, {Pascale}, {Agertz}, {Andersson},
  {Lacchin}, {Lupi}, {Meneghetti}, {Nipoti}, {Ragagnin}, {Rosdahl}, {Vanzella},
  {Vesperini}, \& {Zanella}}]{calura_siege_2025}
{Calura}, F., {Pascale}, R., {Agertz}, O., {et~al.} 2025, \aap, 698, A207,
  \dodoi{10.1051/0004-6361/202452876}

\bibitem[{{Castor} {et~al.}(1975){Castor}, {McCray}, \&
  {Weaver}}]{castor_interstellar_1975}
{Castor}, J., {McCray}, R., \& {Weaver}, R. 1975, \apjl, 200, L107,
  \dodoi{10.1086/181908}

\bibitem[{{Cen}(1992)}]{cen_hydrodynamic_1992}
{Cen}, R. 1992, \apjs, 78, 341, \dodoi{10.1086/191630}

\bibitem[{{Chan} {et~al.}(2019){Chan}, {Kere{\v{s}}}, {Hopkins}, {Quataert},
  {Su}, {Hayward}, \& {Faucher-Gigu{\`e}re}}]{chan_cosmic_2019}
{Chan}, T.~K., {Kere{\v{s}}}, D., {Hopkins}, P.~F., {et~al.} 2019, \mnras, 488,
  3716, \dodoi{10.1093/mnras/stz1895}

\bibitem[{{Chandar} {et~al.}(2017){Chandar}, {Fall}, {Whitmore}, \&
  {Mulia}}]{chandar_fraction_2017}
{Chandar}, R., {Fall}, S.~M., {Whitmore}, B.~C., \& {Mulia}, A.~J. 2017, \apj,
  849, 128, \dodoi{10.3847/1538-4357/aa92ce}

\bibitem[{{Chevance} {et~al.}(2020){Chevance}, {Kruijssen}, {Vazquez-Semadeni},
  {Nakamura}, {Klessen}, {Ballesteros-Paredes}, {Inutsuka}, {Adamo}, \&
  {Hennebelle}}]{chevance_molecular_2020}
{Chevance}, M., {Kruijssen}, J.~M.~D., {Vazquez-Semadeni}, E., {et~al.} 2020,
  \ssr, 216, 50, \dodoi{10.1007/s11214-020-00674-x}

\bibitem[{{Chiu} {et~al.}(2024){Chiu}, {Ruszkowski}, {Thomas}, {Werhahn}, \&
  {Pfrommer}}]{chiu_simulating_2024}
{Chiu}, H. H.~S., {Ruszkowski}, M., {Thomas}, T., {Werhahn}, M., \& {Pfrommer},
  C. 2024, \apj, 976, 136, \dodoi{10.3847/1538-4357/ad84e9}

\bibitem[{{Crocker} {et~al.}(2021){Crocker}, {Krumholz}, \&
  {Thompson}}]{crocker_cosmic_2021}
{Crocker}, R.~M., {Krumholz}, M.~R., \& {Thompson}, T.~A. 2021, \mnras, 502,
  1312, \dodoi{10.1093/mnras/stab148}

\bibitem[{{Dashyan} \& {Dubois}(2020)}]{dashyan_cosmic_2020}
{Dashyan}, G., \& {Dubois}, Y. 2020, \aap, 638, A123,
  \dodoi{10.1051/0004-6361/201936339}

\bibitem[{{Deng} {et~al.}(2024){Deng}, {Li}, {Liu}, {Kannan}, {Smith}, \&
  {Bryan}}]{deng_rigel_2024}
{Deng}, Y., {Li}, H., {Liu}, B., {et~al.} 2024, \aap, 691, A231,
  \dodoi{10.1051/0004-6361/202450699}

\bibitem[{{Draine}(2011)}]{draine_physics_2011}
{Draine}, B.~T. 2011, {Physics of the Interstellar and Intergalactic Medium}
  (Princeton University Press)

\bibitem[{{Farber} {et~al.}(2018){Farber}, {Ruszkowski}, {Yang}, \&
  {Zweibel}}]{farber_impact_2018}
{Farber}, R., {Ruszkowski}, M., {Yang}, H. Y.~K., \& {Zweibel}, E.~G. 2018,
  \apj, 856, 112, \dodoi{10.3847/1538-4357/aab26d}

\bibitem[{{Fensch} {et~al.}(2019){Fensch}, {Duc}, {Boquien}, {Elmegreen},
  {Elmegreen}, {Bournaud}, {Brinks}, {de Grijs}, {Lelli}, {Renaud}, \&
  {Weilbacher}}]{fensch_massive_2019}
{Fensch}, J., {Duc}, P.-A., {Boquien}, M., {et~al.} 2019, \aap, 628, A60,
  \dodoi{10.1051/0004-6361/201834403}

\bibitem[{{Fitz Axen} {et~al.}(2024){Fitz Axen}, {Offner}, {Hopkins},
  {Krumholz}, \& {Grudi{\'c}}}]{fitz_axen_suppressed_2024}
{Fitz Axen}, M., {Offner}, S., {Hopkins}, P.~F., {Krumholz}, M.~R., \&
  {Grudi{\'c}}, M.~Y. 2024, \apj, 973, 16, \dodoi{10.3847/1538-4357/ad675a}

\bibitem[{{Gaia Collaboration} {et~al.}(2016){Gaia Collaboration}, {Prusti},
  {de Bruijne}, {Brown}, {Vallenari}, {Babusiaux}, {Bailer-Jones}, {Bastian},
  {Biermann}, {Evans}, {Eyer}, {Jansen}, {Jordi}, {Klioner}, {Lammers},
  {Lindegren}, {Luri}, {Mignard}, {Milligan}, {Panem}, {Poinsignon},
  {Pourbaix}, {Randich}, {Sarri}, {Sartoretti}, {Siddiqui}, {Soubiran},
  {Valette}, {van Leeuwen}, {Walton}, {Aerts}, {Arenou}, {Cropper}, {Drimmel},
  {H{\o}g}, {Katz}, {Lattanzi}, {O'Mullane}, {Grebel}, {Holland}, {Huc},
  {Passot}, {Bramante}, {Cacciari}, {Casta{\~n}eda}, {Chaoul}, {Cheek}, {De
  Angeli}, {Fabricius}, {Guerra}, {Hern{\'a}ndez}, {Jean-Antoine-Piccolo},
  {Masana}, {Messineo}, {Mowlavi}, {Nienartowicz}, {Ord{\'o}{\~n}ez-Blanco},
  {Panuzzo}, {Portell}, {Richards}, {Riello}, {Seabroke}, {Tanga},
  {Th{\'e}venin}, {Torra}, {Els}, {Gracia-Abril}, {Comoretto},
  {Garcia-Reinaldos}, {Lock}, {Mercier}, {Altmann}, {Andrae}, {Astraatmadja},
  {Bellas-Velidis}, {Benson}, {Berthier}, {Blomme}, {Busso}, {Carry},
  {Cellino}, {Clementini}, {Cowell}, {Creevey}, {Cuypers}, {Davidson}, {De
  Ridder}, {de Torres}, {Delchambre}, {Dell'Oro}, {Ducourant}, {Fr{\'e}mat},
  {Garc{\'\i}a-Torres}, {Gosset}, {Halbwachs}, {Hambly}, {Harrison}, {Hauser},
  {Hestroffer}, {Hodgkin}, {Huckle}, {Hutton}, {Jasniewicz}, {Jordan},
  {Kontizas}, {Korn}, {Lanzafame}, {Manteiga}, {Moitinho}, {Muinonen},
  {Osinde}, {Pancino}, {Pauwels}, {Petit}, {Recio-Blanco}, {Robin}, {Sarro},
  {Siopis}, {Smith}, {Smith}, {Sozzetti}, {Thuillot}, {van Reeven}, {Viala},
  {Abbas}, {Abreu Aramburu}, {Accart}, {Aguado}, {Allan}, {Allasia},
  {Altavilla}, {{\'A}lvarez}, {Alves}, {Anderson}, {Andrei}, {Anglada Varela},
  {Antiche}, {Antoja}, {Ant{\'o}n}, {Arcay}, {Atzei}, {Ayache}, {Bach},
  {Baker}, {Balaguer-N{\'u}{\~n}ez}, {Barache}, {Barata}, {Barbier}, {Barblan},
  {Baroni}, {Barrado y Navascu{\'e}s}, {Barros}, {Barstow}, {Becciani},
  {Bellazzini}, {Bellei}, {Bello Garc{\'\i}a}, {Belokurov}, {Bendjoya},
  {Berihuete}, {Bianchi}, {Bienaym{\'e}}, {Billebaud}, {Blagorodnova},
  {Blanco-Cuaresma}, {Boch}, {Bombrun}, {Borrachero}, {Bouquillon}, {Bourda},
  {Bouy}, {Bragaglia}, {Breddels}, {Brouillet}, {Br{\"u}semeister},
  {Bucciarelli}, {Budnik}, {Burgess}, {Burgon}, {Burlacu}, {Busonero}, {Buzzi},
  {Caffau}, {Cambras}, {Campbell}, {Cancelliere}, {Cantat-Gaudin}, {Carlucci},
  {Carrasco}, {Castellani}, {Charlot}, {Charnas}, {Charvet}, {Chassat},
  {Chiavassa}, {Clotet}, {Cocozza}, {Collins}, {Collins}, \&
  {Costigan}}]{gaia_collaboration_gaia_2016}
{Gaia Collaboration}, {Prusti}, T., {de Bruijne}, J.~H.~J., {et~al.} 2016,
  \aap, 595, A1, \dodoi{10.1051/0004-6361/201629272}

\bibitem[{{Gatto} {et~al.}(2015){Gatto}, {Walch}, {Low}, {Naab}, {Girichidis},
  {Glover}, {W{\"u}nsch}, {Klessen}, {Clark}, {Baczynski}, {Peters},
  {Ostriker}, {Ib{\'a}{\~n}ez-Mej{\'\i}a}, \& {Haid}}]{gatto_modelling_2015}
{Gatto}, A., {Walch}, S., {Low}, M. M.~M., {et~al.} 2015, \mnras, 449, 1057,
  \dodoi{10.1093/mnras/stv324}

\bibitem[{{Gatto} {et~al.}(2017){Gatto}, {Walch}, {Naab}, {Girichidis},
  {W{\"u}nsch}, {Glover}, {Klessen}, {Clark}, {Peters}, {Derigs}, {Baczynski},
  \& {Puls}}]{gatto_silcc_2017}
{Gatto}, A., {Walch}, S., {Naab}, T., {et~al.} 2017, \mnras, 466, 1903,
  \dodoi{10.1093/mnras/stw3209}

\bibitem[{{Ginsburg} \& {Kruijssen}(2018)}]{ginsburg_high_2018}
{Ginsburg}, A., \& {Kruijssen}, J.~M.~D. 2018, \apjl, 864, L17,
  \dodoi{10.3847/2041-8213/aada89}

\bibitem[{{Glover} \& {Clark}(2012)}]{glover_approximations_2012}
{Glover}, S. C.~O., \& {Clark}, P.~C. 2012, \mnras, 421, 116,
  \dodoi{10.1111/j.1365-2966.2011.20260.x}

\bibitem[{{Glover} \& {Jappsen}(2007)}]{glover_star_2007}
{Glover}, S.~C.~O., \& {Jappsen}, A.~K. 2007, \apj, 666, 1,
  \dodoi{10.1086/519445}

\bibitem[{{Goddard} {et~al.}(2010){Goddard}, {Bastian}, \&
  {Kennicutt}}]{goddard_fraction_2010}
{Goddard}, Q.~E., {Bastian}, N., \& {Kennicutt}, R.~C. 2010, \mnras, 405, 857,
  \dodoi{10.1111/j.1365-2966.2010.16511.x}

\bibitem[{{Grassi} {et~al.}(2014){Grassi}, {Bovino}, {Schleicher}, {Prieto},
  {Seifried}, {Simoncini}, \& {Gianturco}}]{grassi_krome_2014}
{Grassi}, T., {Bovino}, S., {Schleicher}, D.~R.~G., {et~al.} 2014, \mnras, 439,
  2386, \dodoi{10.1093/mnras/stu114}

\bibitem[{{Grudi{\'c}} {et~al.}(2021){Grudi{\'c}}, {Guszejnov}, {Hopkins},
  {Offner}, \& {Faucher-Gigu{\`e}re}}]{grudic_starforge_2021}
{Grudi{\'c}}, M.~Y., {Guszejnov}, D., {Hopkins}, P.~F., {Offner}, S. S.~R., \&
  {Faucher-Gigu{\`e}re}, C.-A. 2021, \mnras, 506, 2199,
  \dodoi{10.1093/mnras/stab1347}

\bibitem[{{Grudi{\'c}} {et~al.}(2023){Grudi{\'c}}, {Hafen}, {Rodriguez},
  {Guszejnov}, {Lamberts}, {Wetzel}, {Boylan-Kolchin}, \&
  {Faucher-Gigu{\`e}re}}]{grudic_great_2023}
{Grudi{\'c}}, M.~Y., {Hafen}, Z., {Rodriguez}, C.~L., {et~al.} 2023, \mnras,
  519, 1366, \dodoi{10.1093/mnras/stac3573}

\bibitem[{{Grudi{\'c}} {et~al.}(2018){Grudi{\'c}}, {Hopkins},
  {Faucher-Gigu{\`e}re}, {Quataert}, {Murray}, \&
  {Kere{\v{s}}}}]{grudic_when_2018}
{Grudi{\'c}}, M.~Y., {Hopkins}, P.~F., {Faucher-Gigu{\`e}re}, C.-A., {et~al.}
  2018, \mnras, 475, 3511, \dodoi{10.1093/mnras/sty035}

\bibitem[{{Grudic} {et~al.}(2023){Grudic}, {Offner}, {Guszejnov},
  {Faucher-Gigu{\`e}re}, \& {Hopkins}}]{grudic_does_2023}
{Grudic}, M.~Y., {Offner}, S. S.~R., {Guszejnov}, D., {Faucher-Gigu{\`e}re},
  C.-A., \& {Hopkins}, P.~F. 2023, The Open Journal of Astrophysics, 6, 48,
  \dodoi{10.21105/astro.2307.00052}

\bibitem[{{Gutcke}(2024)}]{gutcke_2024}
{Gutcke}, T.~A. 2024, \apj, 971, 103, \dodoi{10.3847/1538-4357/ad5c62}

\bibitem[{{Gutcke} {et~al.}(2021){Gutcke}, {Pakmor}, {Naab}, \&
  {Springel}}]{gutcke_lyra_2021}
{Gutcke}, T.~A., {Pakmor}, R., {Naab}, T., \& {Springel}, V. 2021, \mnras, 501,
  5597, \dodoi{10.1093/mnras/staa3875}

\bibitem[{{Gutcke} {et~al.}(2022){Gutcke}, {Pfrommer}, {Bryan}, {Pakmor},
  {Springel}, \& {Naab}}]{gutcke_lyra_2022-1}
{Gutcke}, T.~A., {Pfrommer}, C., {Bryan}, G.~L., {et~al.} 2022, \apj, 941, 120,
  \dodoi{10.3847/1538-4357/aca1b4}

\bibitem[{{Hartmann} {et~al.}(2001){Hartmann}, {Ballesteros-Paredes}, \&
  {Bergin}}]{hartmann_rapid_2001}
{Hartmann}, L., {Ballesteros-Paredes}, J., \& {Bergin}, E.~A. 2001, \apj, 562,
  852, \dodoi{10.1086/323863}

\bibitem[{{Hislop} {et~al.}(2022){Hislop}, {Naab}, {Steinwandel}, {Lah{\'e}n},
  {Irodotou}, {Johansson}, \& {Walch}}]{hislop_challenge_2022}
{Hislop}, J.~M., {Naab}, T., {Steinwandel}, U.~P., {et~al.} 2022, \mnras, 509,
  5938, \dodoi{10.1093/mnras/stab3347}

\bibitem[{{Hix} {et~al.}(2025){Hix}, {Armillotta}, {Ostriker}, \&
  {Kim}}]{hix_dynamically_2025}
{Hix}, R.~N., {Armillotta}, L., {Ostriker}, E., \& {Kim}, C.-G. 2025, \apj,
  994, 45, \dodoi{10.3847/1538-4357/ae08b1}

\bibitem[{{Hu} {et~al.}(2016){Hu}, {Naab}, {Walch}, {Glover}, \&
  {Clark}}]{hu_star_2016}
{Hu}, C.-Y., {Naab}, T., {Walch}, S., {Glover}, S. C.~O., \& {Clark}, P.~C.
  2016, \mnras, 458, 3528, \dodoi{10.1093/mnras/stw544}

\bibitem[{{Hu} {et~al.}(2023){Hu}, {Smith}, {Teyssier}, {Bryan}, {Verbeke},
  {Emerick}, {Somerville}, {Burkhart}, {Li}, {Forbes}, \&
  {Starkenburg}}]{hu_code_2023}
{Hu}, C.-Y., {Smith}, M.~C., {Teyssier}, R., {et~al.} 2023, \apj, 950, 132,
  \dodoi{10.3847/1538-4357/accf9e}

\bibitem[{{Jiang} \& {Oh}(2018)}]{jiang_new_2018}
{Jiang}, Y.-F., \& {Oh}, S.~P. 2018, \apj, 854, 5,
  \dodoi{10.3847/1538-4357/aaa6ce}

\bibitem[{{Johnson} {et~al.}(2016){Johnson}, {Seth}, {Dalcanton}, {Beerman},
  {Fouesneau}, {Lewis}, {Weisz}, {Williams}, {Bell}, {Dolphin}, {Larsen},
  {Sandstrom}, \& {Skillman}}]{johnson_panchromatic_2016}
{Johnson}, L.~C., {Seth}, A.~C., {Dalcanton}, J.~J., {et~al.} 2016, \apj, 827,
  33, \dodoi{10.3847/0004-637X/827/1/33}

\bibitem[{{Johnson} {et~al.}(2017){Johnson}, {Seth}, {Dalcanton}, {Beerman},
  {Fouesneau}, {Weisz}, {Bell}, {Dolphin}, {Sandstrom}, \&
  {Williams}}]{johnson_panchromatic_2017}
---. 2017, \apj, 839, 78, \dodoi{10.3847/1538-4357/aa6a1f}

\bibitem[{{Kauffmann} {et~al.}(2013){Kauffmann}, {Pillai}, \&
  {Goldsmith}}]{kauffmann_low_2013}
{Kauffmann}, J., {Pillai}, T., \& {Goldsmith}, P.~F. 2013, \apj, 779, 185,
  \dodoi{10.1088/0004-637X/779/2/185}

\bibitem[{{Keller} {et~al.}(2022){Keller}, {Kruijssen}, \&
  {Chevance}}]{keller_empirically_2022}
{Keller}, B.~W., {Kruijssen}, J.~M.~D., \& {Chevance}, M. 2022, \mnras, 514,
  5355, \dodoi{10.1093/mnras/stac1607}

\bibitem[{{Kim} \& {Ostriker}(2017)}]{kim_three-phase_2017}
{Kim}, C.-G., \& {Ostriker}, E.~C. 2017, \apj, 846, 133,
  \dodoi{10.3847/1538-4357/aa8599}

\bibitem[{{Kim} {et~al.}(2023){Kim}, {Gong}, {Kim}, \&
  {Ostriker}}]{kim_photochemistry_2023}
{Kim}, J.-G., {Gong}, M., {Kim}, C.-G., \& {Ostriker}, E.~C. 2023, \apjs, 264,
  10, \dodoi{10.3847/1538-4365/ac9b1d}

\bibitem[{{Kjellgren} {et~al.}(2025){Kjellgren}, {Girichidis}, {G{\"o}ller},
  {Brucy}, {Klessen}, {Tress}, {Soler}, {Pfrommer}, {Werhahn}, {Glover},
  {Smith}, {Testi}, \& {Molinari}}]{kjellgren_dynamical_2025}
{Kjellgren}, K., {Girichidis}, P., {G{\"o}ller}, J., {et~al.} 2025, \aap, 700,
  A124, \dodoi{10.1051/0004-6361/202553754}

\bibitem[{{Krause} {et~al.}(2020){Krause}, {Offner}, {Charbonnel}, {Gieles},
  {Klessen}, {V{\'a}zquez-Semadeni}, {Ballesteros-Paredes}, {Girichidis},
  {Kruijssen}, {Ward}, \& {Zinnecker}}]{krause_physics_2020}
{Krause}, M. G.~H., {Offner}, S. S.~R., {Charbonnel}, C., {et~al.} 2020, \ssr,
  216, 64, \dodoi{10.1007/s11214-020-00689-4}

\bibitem[{{Kravtsov}(2003)}]{kravtsov_origin_2003}
{Kravtsov}, A.~V. 2003, \apjl, 590, L1, \dodoi{10.1086/376674}

\bibitem[{{Kroupa}(1995)}]{kroupa_inverse_1995}
{Kroupa}, P. 1995, \mnras, 277, 1491, \dodoi{10.1093/mnras/277.4.1491}

\bibitem[{{Kruijssen}(2012)}]{kruijssen_fraction_2012}
{Kruijssen}, J.~M.~D. 2012, \mnras, 426, 3008,
  \dodoi{10.1111/j.1365-2966.2012.21923.x}

\bibitem[{{Krumholz} {et~al.}(2019){Krumholz}, {McKee}, \&
  {Bland-Hawthorn}}]{krumholz_star_2019}
{Krumholz}, M.~R., {McKee}, C.~F., \& {Bland-Hawthorn}, J. 2019, \araa, 57,
  227, \dodoi{10.1146/annurev-astro-091918-104430}

\bibitem[{{Kulsrud} \& {Pearce}(1969)}]{kulsrud_effect_1969}
{Kulsrud}, R., \& {Pearce}, W.~P. 1969, \apj, 156, 445, \dodoi{10.1086/149981}

\bibitem[{{Lada} \& {Lada}(2003)}]{lada_embedded_2003}
{Lada}, C.~J., \& {Lada}, E.~A. 2003, \araa, 41, 57,
  \dodoi{10.1146/annurev.astro.41.011802.094844}

\bibitem[{{Lah{\'e}n} {et~al.}(2020){Lah{\'e}n}, {Naab}, {Johansson},
  {Elmegreen}, {Hu}, {Walch}, {Steinwandel}, \& {Moster}}]{lahen_griffin_2020}
{Lah{\'e}n}, N., {Naab}, T., {Johansson}, P.~H., {et~al.} 2020, \apj, 891, 2,
  \dodoi{10.3847/1538-4357/ab7190}

\bibitem[{{Larsen}(2002)}]{larsen_luminosity_2002}
{Larsen}, S.~S. 2002, \aj, 124, 1393, \dodoi{10.1086/342381}

\bibitem[{{Larson}(1981)}]{larson_turbulence_1981}
{Larson}, R.~B. 1981, \mnras, 194, 809, \dodoi{10.1093/mnras/194.4.809}

\bibitem[{{Lemmerz} {et~al.}(2025){Lemmerz}, {Shalaby}, {Pfrommer}, \&
  {Thomas}}]{lemmerz_theory_2025}
{Lemmerz}, R., {Shalaby}, M., {Pfrommer}, C., \& {Thomas}, T. 2025, \apj, 979,
  34, \dodoi{10.3847/1538-4357/ad8eb3}

\bibitem[{{Li} \& {Gnedin}(2019)}]{li_star_2019}
{Li}, H., \& {Gnedin}, O.~Y. 2019, \mnras, 486, 4030,
  \dodoi{10.1093/mnras/stz1114}

\bibitem[{{Li} {et~al.}(2018){Li}, {Gnedin}, \& {Gnedin}}]{li_star_2018}
{Li}, H., {Gnedin}, O.~Y., \& {Gnedin}, N.~Y. 2018, \apj, 861, 107,
  \dodoi{10.3847/1538-4357/aac9b8}

\bibitem[{{Li} {et~al.}(2017){Li}, {Gnedin}, {Gnedin}, {Meng}, {Semenov}, \&
  {Kravtsov}}]{li_star_2017}
{Li}, H., {Gnedin}, O.~Y., {Gnedin}, N.~Y., {et~al.} 2017, \apj, 834, 69,
  \dodoi{10.3847/1538-4357/834/1/69}

\bibitem[{{Li} {et~al.}(2019){Li}, {Vogelsberger}, {Marinacci}, \&
  {Gnedin}}]{li_disruption_2019}
{Li}, H., {Vogelsberger}, M., {Marinacci}, F., \& {Gnedin}, O.~Y. 2019, \mnras,
  487, 364, \dodoi{10.1093/mnras/stz1271}

\bibitem[{{Li} {et~al.}(2020){Li}, {Vogelsberger}, {Marinacci}, {Sales}, \&
  {Torrey}}]{li_effects_2020}
{Li}, H., {Vogelsberger}, M., {Marinacci}, F., {Sales}, L.~V., \& {Torrey}, P.
  2020, \mnras, 499, 5862, \dodoi{10.1093/mnras/staa3122}

\bibitem[{{Lim} \& {Lee}(2015)}]{lim_star_2015}
{Lim}, S., \& {Lee}, M.~G. 2015, \apj, 804, 123,
  \dodoi{10.1088/0004-637X/804/2/123}

\bibitem[{{Longmore} {et~al.}(2014){Longmore}, {Kruijssen}, {Bastian}, {Bally},
  {Rathborne}, {Testi}, {Stolte}, {Dale}, {Bressert}, \&
  {Alves}}]{longmore_formation_2014}
{Longmore}, S.~N., {Kruijssen}, J.~M.~D., {Bastian}, N., {et~al.} 2014, in
  Protostars and Planets VI, ed. H.~{Beuther}, R.~S. {Klessen}, C.~P.
  {Dullemond}, \& T.~{Henning}, 291--314,
  \dodoi{10.2458/azu_uapress_9780816531240-ch013}

\bibitem[{{McCray} \& {Kafatos}(1987)}]{mccray_supershells_1987}
{McCray}, R., \& {Kafatos}, M. 1987, \apj, 317, 190, \dodoi{10.1086/165267}

\bibitem[{{McKee} {et~al.}(1984){McKee}, {van Buren}, \&
  {Lazareff}}]{mckee_photoionized_1984}
{McKee}, C.~F., {van Buren}, D., \& {Lazareff}, B. 1984, \apjl, 278, L115,
  \dodoi{10.1086/184237}

\bibitem[{{Messa} {et~al.}(2018){Messa}, {Adamo}, {Calzetti}, {Reina-Campos},
  {Colombo}, {Schinnerer}, {Chandar}, {Dale}, {Gouliermis}, {Grasha}, {Grebel},
  {Elmegreen}, {Fumagalli}, {Johnson}, {Kruijssen}, {{\"O}stlin}, {Shabani},
  {Smith}, \& {Whitmore}}]{messa_young_2018}
{Messa}, M., {Adamo}, A., {Calzetti}, D., {et~al.} 2018, \mnras, 477, 1683,
  \dodoi{10.1093/mnras/sty577}

\bibitem[{{Moseley} {et~al.}(2021){Moseley}, {Draine}, {Tomida}, \&
  {Stone}}]{moseley_turbulent_2021}
{Moseley}, E.~R., {Draine}, B.~T., {Tomida}, K., \& {Stone}, J.~M. 2021,
  \mnras, 500, 3290, \dodoi{10.1093/mnras/staa3384}

\bibitem[{{Myers} {et~al.}(2025){Myers}, {Heyer}, {Stephens}, {Coud{\'e}},
  {Karnath}, \& {Smith}}]{myers_gravitational_2025}
{Myers}, P.~C., {Heyer}, M., {Stephens}, I.~W., {et~al.} 2025, arXiv e-prints,
  arXiv:2508.05826, \dodoi{10.48550/arXiv.2508.05826}

\bibitem[{{Naab} \& {Ostriker}(2017)}]{naab_theoretical_2017}
{Naab}, T., \& {Ostriker}, J.~P. 2017, \araa, 55, 59,
  \dodoi{10.1146/annurev-astro-081913-040019}

\bibitem[{{Ni} {et~al.}(2025){Ni}, {Li}, {Vogelsberger}, {Sales}, {Marinacci},
  \& {Torrey}}]{ni_life_2025}
{Ni}, Y., {Li}, H., {Vogelsberger}, M., {et~al.} 2025, \aap, 699, A282,
  \dodoi{10.1051/0004-6361/202554126}

\bibitem[{{Oka}(2006)}]{oka_interstellar_2006}
{Oka}, T. 2006, Proceedings of the National Academy of Science, 103, 12235,
  \dodoi{10.1073/pnas.0601242103}

\bibitem[{{Oort} \& {Spitzer}(1955)}]{oort_acceleration_1955}
{Oort}, J.~H., \& {Spitzer}, Jr., L. 1955, \apj, 121, 6, \dodoi{10.1086/145958}

\bibitem[{{Owen}(2023)}]{owen_secret_2023}
{Owen}, E. 2023, Astronomy and Geophysics, 64, 1.29,
  \dodoi{10.1093/astrogeo/atac090}

\bibitem[{{Owen} {et~al.}(2023){Owen}, {Wu}, {Inoue}, {Yang}, \&
  {Mitchell}}]{owen_cosmic_2023}
{Owen}, E.~R., {Wu}, K., {Inoue}, Y., {Yang}, H. Y.~K., \& {Mitchell}, A. M.~W.
  2023, Galaxies, 11, 86, \dodoi{10.3390/galaxies11040086}

\bibitem[{{Padovani} {et~al.}(2009){Padovani}, {Galli}, \&
  {Glassgold}}]{padovani_cosmic-ray_2009}
{Padovani}, M., {Galli}, D., \& {Glassgold}, A.~E. 2009, \aap, 501, 619,
  \dodoi{10.1051/0004-6361/200911794}

\bibitem[{{Padovani} {et~al.}(2020){Padovani}, {Ivlev}, {Galli}, {Offner},
  {Indriolo}, {Rodgers-Lee}, {Marcowith}, {Girichidis}, {Bykov}, \&
  {Kruijssen}}]{padovani_impact_2020}
{Padovani}, M., {Ivlev}, A.~V., {Galli}, D., {et~al.} 2020, \ssr, 216, 29,
  \dodoi{10.1007/s11214-020-00654-1}

\bibitem[{{Pakmor} {et~al.}(2011){Pakmor}, {Bauer}, \&
  {Springel}}]{pakmor_magnetohydrodynamics_2011}
{Pakmor}, R., {Bauer}, A., \& {Springel}, V. 2011, \mnras, 418, 1392,
  \dodoi{10.1111/j.1365-2966.2011.19591.x}

\bibitem[{{Pakmor} {et~al.}(2016{\natexlab{a}}){Pakmor}, {Pfrommer}, {Simpson},
  \& {Springel}}]{pakmor_galactic_2016}
{Pakmor}, R., {Pfrommer}, C., {Simpson}, C.~M., \& {Springel}, V.
  2016{\natexlab{a}}, \apjl, 824, L30, \dodoi{10.3847/2041-8205/824/2/L30}

\bibitem[{{Pakmor} \& {Springel}(2013)}]{pakmor_simulations_2013}
{Pakmor}, R., \& {Springel}, V. 2013, \mnras, 432, 176,
  \dodoi{10.1093/mnras/stt428}

\bibitem[{{Pakmor} {et~al.}(2016{\natexlab{b}}){Pakmor}, {Springel}, {Bauer},
  {Mocz}, {Munoz}, {Ohlmann}, {Schaal}, \& {Zhu}}]{pakmor_improving_2016}
{Pakmor}, R., {Springel}, V., {Bauer}, A., {et~al.} 2016{\natexlab{b}}, \mnras,
  455, 1134, \dodoi{10.1093/mnras/stv2380}

\bibitem[{{Pasquali} {et~al.}(2011){Pasquali}, {Bik}, {Zibetti}, {Ageorges},
  {Seifert}, {Brandner}, {Rix}, {J{\"u}tte}, {Knierim}, {Buschkamp}, {Feiz},
  {Gemperlein}, {Germeroth}, {Hofmann}, {Laun}, {Lederer}, {Lehmitz}, {Lenzen},
  {Mall}, {Mandel}, {M{\"u}ller}, {Naranjo}, {Polsterer}, {Quirrenbach},
  {Sch{\"a}ffner}, {Storz}, \& {Weiser}}]{pasquali_infrared_2011}
{Pasquali}, A., {Bik}, A., {Zibetti}, S., {et~al.} 2011, \aj, 141, 132,
  \dodoi{10.1088/0004-6256/141/4/132}

\bibitem[{{Pfrommer} {et~al.}(2017){Pfrommer}, {Pakmor}, {Schaal}, {Simpson},
  \& {Springel}}]{pfrommer_simulating_2017}
{Pfrommer}, C., {Pakmor}, R., {Schaal}, K., {Simpson}, C.~M., \& {Springel}, V.
  2017, \mnras, 465, 4500, \dodoi{10.1093/mnras/stw2941}

\bibitem[{{Rafelski} \& {Zaritsky}(2005)}]{rafelski_star_2005}
{Rafelski}, M., \& {Zaritsky}, D. 2005, \aj, 129, 2701, \dodoi{10.1086/424938}

\bibitem[{{Rathjen} {et~al.}(2023){Rathjen}, {Naab}, {Walch}, {Seifried},
  {Girichidis}, \& {W{\"u}nsch}}]{rathjen_silcc_2023}
{Rathjen}, T.-E., {Naab}, T., {Walch}, S., {et~al.} 2023, \mnras, 522, 1843,
  \dodoi{10.1093/mnras/stad1104}

\bibitem[{{Rathjen} {et~al.}(2021){Rathjen}, {Naab}, {Girichidis}, {Walch},
  {W{\"u}nsch}, {Dinnbier}, {Seifried}, {Klessen}, \&
  {Glover}}]{rathjen_silcc_2021}
{Rathjen}, T.-E., {Naab}, T., {Girichidis}, P., {et~al.} 2021, \mnras, 504,
  1039, \dodoi{10.1093/mnras/stab900}

\bibitem[{{Reina-Campos} {et~al.}(2025){Reina-Campos}, {Gnedin}, {Sills}, \&
  {Li}}]{reina-campos_star_2025}
{Reina-Campos}, M., {Gnedin}, O.~Y., {Sills}, A., \& {Li}, H. 2025, \apj, 978,
  15, \dodoi{10.3847/1538-4357/ad909f}

\bibitem[{{Reina-Campos} \& {Kruijssen}(2017)}]{reina-campos_unified_2017}
{Reina-Campos}, M., \& {Kruijssen}, J.~M.~D. 2017, \mnras, 469, 1282,
  \dodoi{10.1093/mnras/stx790}

\bibitem[{{Rogers} \& {Pittard}(2013)}]{rogers_feedback_2013}
{Rogers}, H., \& {Pittard}, J.~M. 2013, \mnras, 431, 1337,
  \dodoi{10.1093/mnras/stt255}

\bibitem[{{Ruszkowski} \& {Pfrommer}(2023)}]{ruszkowski_cosmic_2023}
{Ruszkowski}, M., \& {Pfrommer}, C. 2023, \aapr, 31, 4,
  \dodoi{10.1007/s00159-023-00149-2}

\bibitem[{{Ruszkowski} {et~al.}(2017){Ruszkowski}, {Yang}, \&
  {Zweibel}}]{ruszkowski_global_2017}
{Ruszkowski}, M., {Yang}, H. Y.~K., \& {Zweibel}, E. 2017, \apj, 834, 208,
  \dodoi{10.3847/1538-4357/834/2/208}

\bibitem[{{Ryon} {et~al.}(2014){Ryon}, {Adamo}, {Bastian}, {Smith},
  {Gallagher}, {Konstantopoulos}, {Larsen}, {Silva-Villa}, \&
  {Zackrisson}}]{ryon_snapshot_2014}
{Ryon}, J.~E., {Adamo}, A., {Bastian}, N., {et~al.} 2014, \aj, 148, 33,
  \dodoi{10.1088/0004-6256/148/2/33}

\bibitem[{{Salem} \& {Bryan}(2014)}]{salem_cosmic_2014}
{Salem}, M., \& {Bryan}, G.~L. 2014, \mnras, 437, 3312,
  \dodoi{10.1093/mnras/stt2121}

\bibitem[{{Schmidt}(1959)}]{schmidt_rate_1959}
{Schmidt}, M. 1959, \apj, 129, 243, \dodoi{10.1086/146614}

\bibitem[{{Shalaby} {et~al.}(2023){Shalaby}, {Thomas}, {Pfrommer}, {Lemmerz},
  \& {Bresci}}]{shalaby_deciphering_2023}
{Shalaby}, M., {Thomas}, T., {Pfrommer}, C., {Lemmerz}, R., \& {Bresci}, V.
  2023, Journal of Plasma Physics, 89, 175890603,
  \dodoi{10.1017/S0022377823001289}

\bibitem[{{Sike} {et~al.}(2025){Sike}, {Thomas}, {Ruszkowski}, {Pfrommer}, \&
  {Weber}}]{sike_cosmic-ray-driven_2025}
{Sike}, B., {Thomas}, T., {Ruszkowski}, M., {Pfrommer}, C., \& {Weber}, M.
  2025, \apj, 987, 204, \dodoi{10.3847/1538-4357/adda3d}

\bibitem[{{Silva-Villa} \& {Larsen}(2011)}]{silva-villa_star_2011}
{Silva-Villa}, E., \& {Larsen}, S.~S. 2011, \aap, 529, A25,
  \dodoi{10.1051/0004-6361/201016206}

\bibitem[{{Simpson} {et~al.}(2016){Simpson}, {Pakmor}, {Marinacci}, {Pfrommer},
  {Springel}, {Glover}, {Clark}, \& {Smith}}]{simpson_role_2016}
{Simpson}, C.~M., {Pakmor}, R., {Marinacci}, F., {et~al.} 2016, \apjl, 827,
  L29, \dodoi{10.3847/2041-8205/827/2/L29}

\bibitem[{Simpson(1951)}]{simpson_interpretation_1951}
Simpson, E.~H. 1951, Journal of the Royal Statistical Society: Series B
  (Methodological), 13, 238, \dodoi{10.1111/j.2517-6161.1951.tb00088.x}

\bibitem[{{Skilling}(1975)}]{skilling_cosmic_1975}
{Skilling}, J. 1975, \mnras, 172, 557, \dodoi{10.1093/mnras/172.3.557}

\bibitem[{{Smith}(2021)}]{smith_sensitivity_2021}
{Smith}, M.~C. 2021, \mnras, 502, 5417, \dodoi{10.1093/mnras/stab291}

\bibitem[{{Smith} {et~al.}(2021){Smith}, {Bryan}, {Somerville}, {Hu},
  {Teyssier}, {Burkhart}, \& {Hernquist}}]{smith_efficient_2021}
{Smith}, M.~C., {Bryan}, G.~L., {Somerville}, R.~S., {et~al.} 2021, \mnras,
  506, 3882, \dodoi{10.1093/mnras/stab1896}

\bibitem[{{Socrates} {et~al.}(2008){Socrates}, {Davis}, \&
  {Ramirez-Ruiz}}]{socrates_eddington_2008}
{Socrates}, A., {Davis}, S.~W., \& {Ramirez-Ruiz}, E. 2008, \apj, 687, 202,
  \dodoi{10.1086/590046}

\bibitem[{{Springel}(2010)}]{springel_e_2010}
{Springel}, V. 2010, \mnras, 401, 791, \dodoi{10.1111/j.1365-2966.2009.15715.x}

\bibitem[{{Springel} \& {Hernquist}(2003)}]{springel_cosmological_2003}
{Springel}, V., \& {Hernquist}, L. 2003, \mnras, 339, 289,
  \dodoi{10.1046/j.1365-8711.2003.06206.x}

\bibitem[{{Str{\"o}mgren}(1939)}]{stromgren_physical_1939}
{Str{\"o}mgren}, B. 1939, \apj, 89, 526, \dodoi{10.1086/144074}

\bibitem[{{Thomas} \& {Pfrommer}(2019)}]{thomas_cosmic-ray_2019}
{Thomas}, T., \& {Pfrommer}, C. 2019, \mnras, 485, 2977,
  \dodoi{10.1093/mnras/stz263}

\bibitem[{{Thomas} \& {Pfrommer}(2022)}]{thomas_comparing_2022}
---. 2022, \mnras, 509, 4803, \dodoi{10.1093/mnras/stab3079}

\bibitem[{{Thomas} {et~al.}(2021){Thomas}, {Pfrommer}, \&
  {Pakmor}}]{thomas_finite_2021}
{Thomas}, T., {Pfrommer}, C., \& {Pakmor}, R. 2021, \mnras, 503, 2242,
  \dodoi{10.1093/mnras/stab397}

\bibitem[{{Thomas} {et~al.}(2023){Thomas}, {Pfrommer}, \&
  {Pakmor}}]{thomas_cosmic-ray-driven_2023}
---. 2023, \mnras, 521, 3023, \dodoi{10.1093/mnras/stad472}

\bibitem[{{Thomas} {et~al.}(2025{\natexlab{a}}){Thomas}, {Pfrommer}, \&
  {Pakmor}}]{thomas_why_2025}
---. 2025{\natexlab{a}}, \aap, 698, A104, \dodoi{10.1051/0004-6361/202450817}

\bibitem[{{Thomas} {et~al.}(2025{\natexlab{b}}){Thomas}, {Pfrommer}, {Pakmor},
  {Lemmerz}, \& {Shalaby}}]{thomas_effective_2025}
{Thomas}, T., {Pfrommer}, C., {Pakmor}, R., {Lemmerz}, R., \& {Shalaby}, M.
  2025{\natexlab{b}}, arXiv e-prints, arXiv:2510.16125,
  \dodoi{10.48550/arXiv.2510.16125}

\bibitem[{{Toomre}(1964)}]{toomre_gravitational_1964}
{Toomre}, A. 1964, \apj, 139, 1217, \dodoi{10.1086/147861}

\bibitem[{{Trumpler}(1930)}]{trumpler_preliminary_1930}
{Trumpler}, R.~J. 1930, Lick Observatory Bulletin, 420, 154,
  \dodoi{10.5479/ADS/bib/1930LicOB.14.154T}

\bibitem[{{Uhlig} {et~al.}(2012){Uhlig}, {Pfrommer}, {Sharma}, {Nath},
  {En{\ss}lin}, \& {Springel}}]{uhlig_galactic_2012}
{Uhlig}, M., {Pfrommer}, C., {Sharma}, M., {et~al.} 2012, \mnras, 423, 2374,
  \dodoi{10.1111/j.1365-2966.2012.21045.x}

\bibitem[{{V{\'a}zquez-Semadeni} {et~al.}(2019){V{\'a}zquez-Semadeni}, {Palau},
  {Ballesteros-Paredes}, {G{\'o}mez}, \&
  {Zamora-Avil{\'e}s}}]{vazquez-semadeni_global_2019}
{V{\'a}zquez-Semadeni}, E., {Palau}, A., {Ballesteros-Paredes}, J.,
  {G{\'o}mez}, G.~C., \& {Zamora-Avil{\'e}s}, M. 2019, \mnras, 490, 3061,
  \dodoi{10.1093/mnras/stz2736}

\bibitem[{{Walch} {et~al.}(2015){Walch}, {Girichidis}, {Naab}, {Gatto},
  {Glover}, {W{\"u}nsch}, {Klessen}, {Clark}, {Peters}, {Derigs}, \&
  {Baczynski}}]{walch_silcc_2015}
{Walch}, S., {Girichidis}, P., {Naab}, T., {et~al.} 2015, \mnras, 454, 238,
  \dodoi{10.1093/mnras/stv1975}

\bibitem[{{Weber} {et~al.}(2025){Weber}, {Thomas}, {Pfrommer}, \&
  {Pakmor}}]{weber_crexit_2025}
{Weber}, M., {Thomas}, T., {Pfrommer}, C., \& {Pakmor}, R. 2025, \aap, 698,
  A125, \dodoi{10.1051/0004-6361/202553954}

\bibitem[{{Weinberger} {et~al.}(2020){Weinberger}, {Springel}, \&
  {Pakmor}}]{weinberger_arepo_2020}
{Weinberger}, R., {Springel}, V., \& {Pakmor}, R. 2020, \apjs, 248, 32,
  \dodoi{10.3847/1538-4365/ab908c}

\bibitem[{{Wetzel} {et~al.}(2025){Wetzel}, {Samuel}, {Gandhi}, {Ponnada}, {Su},
  {Arora}, {Angles-Alcazar}, {Hayward}, {Sanderson}, {Feldmann}, {Cochrane},
  {Nikakhtar}, {Panithanpaisal}, {Benavides}, {Pandya}, {Grudic}, {Hummels},
  {Gurvich}, {Hafen}, {Ma}, {Garrison-Kimmel}, {Sameie}, {Chan}, {El-Badry},
  {Necib}, {Loebman}, {Wellons}, {Robles}, {Wheeler}, {Moreno}, {Stern},
  {Boylan-Kolchin}, {Bullock}, {Faucher-Giguere}, {Keres}, {Quataert}, \&
  {Hopkins}}]{wetzel_second_2025}
{Wetzel}, A., {Samuel}, J., {Gandhi}, P.~J., {et~al.} 2025, arXiv e-prints,
  arXiv:2508.06608, \dodoi{10.48550/arXiv.2508.06608}

\bibitem[{{Wheeler} {et~al.}(2019){Wheeler}, {Hopkins}, {Pace},
  {Garrison-Kimmel}, {Boylan-Kolchin}, {Wetzel}, {Bullock}, {Kere{\v{s}}},
  {Faucher-Gigu{\`e}re}, \& {Quataert}}]{wheeler_be_2019}
{Wheeler}, C., {Hopkins}, P.~F., {Pace}, A.~B., {et~al.} 2019, \mnras, 490,
  4447, \dodoi{10.1093/mnras/stz2887}

\bibitem[{{Whitworth}(1979)}]{whitworth_erosion_1979}
{Whitworth}, A. 1979, \mnras, 186, 59, \dodoi{10.1093/mnras/186.1.59}

\bibitem[{{Zhang} {et~al.}(2025){Zhang}, {Sales}, {Gutcke}, {Deng}, {Li},
  {Pakmor}, {Marinacci}, {Springel}, {Vogelsberger}, {Torrey}, {Liu}, {Kannan},
  {Smith}, \& {Bryan}}]{zhang_entangled_2025}
{Zhang}, E., {Sales}, L.~V., {Gutcke}, T.~A., {et~al.} 2025, arXiv e-prints,
  arXiv:2510.02432.
\newblock \doarXiv{2510.02432}

\bibitem[{{Zweibel}(2017)}]{zweibel_basis_2017}
{Zweibel}, E.~G. 2017, Physics of Plasmas, 24, 055402,
  \dodoi{10.1063/1.4984017}

\end{thebibliography}
\bibliographystyle{aasjournal}

\end{document}